\renewcommand{\textbf}[1]{#1}
\documentclass[journal=jacsat,manuscript=article, layout=two-column]{achemso}
\setkeys{acs}{articletitle = true}

\makeatletter
\let\l@addto@macro\relax
\makeatother
\usepackage[fontsize=10pt]{scrextend}
\SectionNumbersOn

\usepackage[version=3]{mhchem} % Formula subscripts using \ce{}

\newcommand\kcalmol{kcal mol$^{-1}$~}

\newcounter{chem}
\newcounter{temp}

\usepackage{multirow}
\usepackage{xcolor}
\usepackage{amssymb}
\usepackage{pifont}
\usepackage{hyperref}
\usepackage{threeparttable}
\usepackage{placeins}
\author{Juan Carlos del Valle}
\affiliation{Institute for Theoretical Chemistry, University of Stuttgart, Pfaffenwaldring 55, D-70569 Stuttgart, Germany}

\author{Miguel Sanz-Novo}
\affiliation{Centro de Astrobiolog{\'i}a (CAB), CSIC-INTA, Carretera de Ajalvir km 4, Torrej{\'o}n de Ardoz, 28850 Madrid, Spain}

\author{Johannes Kästner}
\affiliation{Institute for Theoretical Chemistry, University of Stuttgart, Pfaffenwaldring 55, D-70569 Stuttgart, Germany}

\author{Kenji Furuya}
\affiliation{RIKEN Cluster for Pioneering Research, 2-1 Hirosawa, Wako-shi, Saitama 351-0198, Japan}

\author{Víctor M. Rivilla}
\affiliation{Centro de Astrobiolog{\'i}a (CAB), INTA-CSIC, Carretera de Ajalvir km 4, Torrej{\'o}n de Ardoz, 28850 Madrid, Spain}

\author{Rafael Mart\'in-Dom\'enech}
\affiliation{Centro de Astrobiolog{\'i}a (CAB), INTA-CSIC, Carretera de Ajalvir km 4, Torrej{\'o}n de Ardoz, 28850 Madrid, Spain}

\author{Germán Molpeceres}
\affiliation{Departamento de Astrofísica Molecular, Instituto de Física Fundamental (IFF-CSIC), C/ Serrano 121, E-28006 Madrid, Spain}
\email{german.molpeceres@iff.csic.es}

\title[HNSO]
  {Atom Addition Formation of Thionylimide (HNSO) on Interstellar Dust Grains: Chemical routes requiring oxygen and nitrogen atom surface diffusion}

\abbreviations{IR,NMR,UV}

\begin{document}

%%%%%%%%%%%%%%%%%%%%%%%%%%%%%%%%%%%%%%%%%%%%%%%%%%%%%%%%%%%%%%%%%%%%%
%% The "tocentry" environment can be used to create an entry for the
%% graphical table of contents. It is given here as some journals
%% require that it is printed as part of the abstract page. It will
%% be automatically moved as appropriate.
%%%%%%%%%%%%%%%%%%%%%%%%%%%%%%%%%%%%%%%%%%%%%%%%%%%%%%%%%%%%%%%%%%%%%
% Put this near the end of your document (achemso prints it on a separate page)
\begin{tocentry}
  \centering
  \includegraphics[width=9cm,height=3.5cm,keepaspectratio]{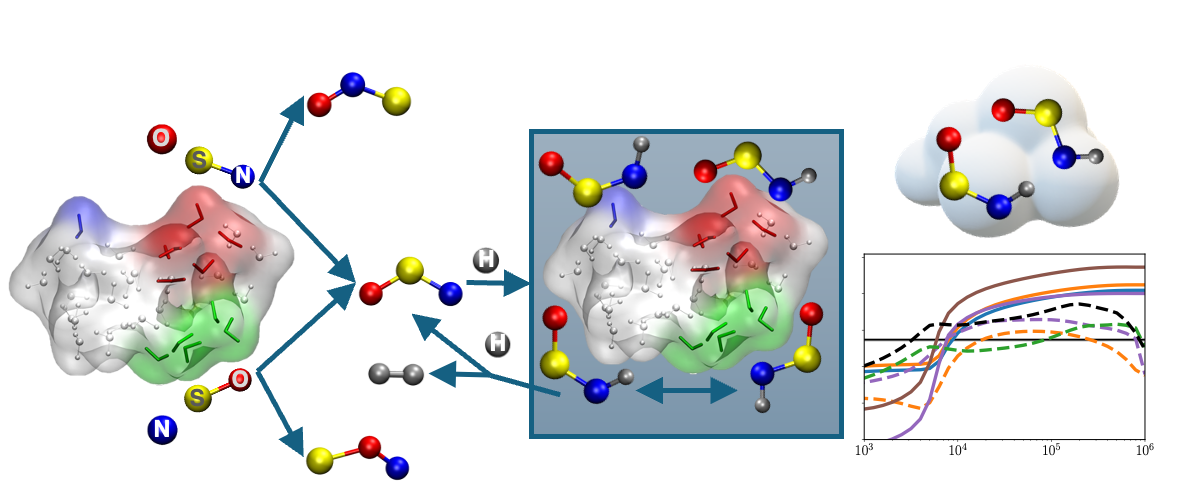}%
  % Optional single-line caption inside the box:
  % \par\vspace{0.2em}\parbox{9cm}{\centering Short descriptive TOC text.}
\end{tocentry}

%%%%%%%%%%%%%%%%%%%%%%%%%%%%%%%%%%%%%%%%%%%%%%%%%%%%%%%%%%%%%%%%%%%%%
%% The abstract environment will automatically gobble the contents
%% if an abstract is not used by the target journal.
%%%%%%%%%%%%%%%%%%%%%%%%%%%%%%%%%%%%%%%%%%%%%%%%%%%%%%%%%%%%%%%%%%%%%
\begin{abstract}
We investigate the formation of the recently detected HNSO molecule using quantum chemical calculations on ices and astrochemical models in tandem. Our results indicate that HNSO is efficiently produced on grain surfaces through reactions involving atomic oxygen and nitrogen atoms with the radicals \ce{NS} and \ce{SO}, forming \ce{NSO} as a key intermediate. Subsequent hydrogenation of \ce{NSO} leads to HNSO, with a clear preference for the lowest energy \textit{cis} conformer, while the \textit{trans} form is metastable and may be short-lived under typical interstellar conditions. The models predict that solid HNSO can reach abundances comparable to icy OCS, placing it among the major sulfur-bearing species in interstellar ices. Gas-phase abundances, in contrast, remain lower than those of OCS. The implementation of a multibinding scheme in our models clarifies the role of diffusive chemistry in the production of HNSO at early times, improving agreement with observations. These findings suggest that reactions involving diffusing O and N atoms on icy grains contribute significantly to sulfur chemistry and beyond in dense clouds and motivate further searches for molecules containing simultaneously H, N, O and S in other astronomical environments.
\end{abstract}

%%%%%%%%%%%%%%%%%%%%%%%%%%%%%%%%%%%%%%%%%%%%%%%%%%%%%%%%%%%%%%%%%%%%%
%% Start the main part of the manuscript here.
%%%%%%%%%%%%%%%%%%%%%%%%%%%%%%%%%%%%%%%%%%%%%%%%%%%%%%%%%%%%%%%%%%%%%
\section{Introduction} \label{sec:intro}

One of the main points of consensus that the astrochemical community needs to reach amid the current surge of new interstellar molecule detections is the definition of what constitutes a complex interstellar molecule. In 2009, and referring specifically to complex organic molecules (COMs), \citet{Herbst2009} defined them as species containing at least \textbf{one carbon atom, and more than five atoms in total. This definition encompasses a wide range of chemical compounds. like PAHs, or prebiotic compounds} \textbf{However, in order to undestand the chemical complexity in the interstellar medium (ISM) it is necessary to understand the formation of molecules that do not strictly fit this definition.}

The molecule subject of this study, thionylimide (HNSO), is certainly not organic, as it lacks a carbon atom in its structure, nor is it complex since it contains only four atoms in total. Nevertheless, it displays a remarkable diversity of heteroatoms, simultaneously harboring N, S and O, three key elements for the development of life, a rare occurrence in interstellar chemical inventories. HNSO was recently discovered toward the Galactic Center molecular cloud G+0.693–0.027 (hereafter G+0.693) \cite{sanz-novo_discovery_2024} as part of an ultra-deep molecular line survey conducted with the Yebes 40 m and IRAM 30 m radio telescopes \cite{Rivilla23,SanzNovo23}. Interestingly, G+0.693 hosts one of the richest chemical inventories in our Galaxy. Its chemistry is governed by large-scale low-velocity shocks, which facilitate the release of material from the icy mantles of dust grains, thereby providing access to grain-surface chemistry \cite{requena-torres2006,Zeng2018}. This makes G+0.693 a unique astrochemical niche for the discovery of new interstellar species \cite{Jimenez-Serra22,SanzNovo23,Rivilla23,Sanz-Novo2025b}. The detection of HNSO opened the door to a completely new family of interstellar chemistry, as it is the first interstellar molecule ever detected containing the NSO moiety, thus representing a promising link between the chemistry of these three elements in space. Within Earth’s biosphere, NSO chemistry plays a key role in cellular and intercellular signaling processes, linking the biochemistries of two essential biological messengers, nitric oxide (NO) and hydrogen sulfide (H$_2$S) \cite{FOSTER2009,Filipovic2012,Miljkovic2013,Ivanova2014,Kumar2017}. NO, recognized as the first gasotransmitter, is involved in regulating vascular tone and cardiac function, while H$_2$S contributes to antioxidative stress defense and inflammation control \cite{Wu2018,Zhao2024}. Moreover, NSO-bearing compounds hold valuable geological and geochemical information \cite{Chang23nso}, effectively recording biotic and palaeoenvironmental signatures,\cite{yue2023} reinforcing their relevance in astrobiological research even further.

In light of the discussion above and the growing evidence that grain processes are viable initiators of chemical complexity in G+0.693, it is natural to examine non-hydrogenative grain chemistry alongside the well-established hydrogenation pathways.\cite{Watanabe2002, Fuchs2009, Rimola2014, molpeceres_carbon_2024,theule_hydrogenation_2011, haupa_hydrogen_2019, gobi_thionethiol_2025,molpeceres_hydrogen_2022, Molpeceres2021b,molpeceres_hydrogenation_2025} Hydrogen additions are favored because H atoms possess unique properties, such as rapid diffusion on amorphous solid water (ASW) surfaces\cite{Hama2012,asg17, SENEVIRATHNE201759} and the ability to tunnel through activation barriers. However, hydrogenation alone cannot account for the formation of HNSO, since its immediate precursor, NSO, contains no hydrogen. Moreover, the NSO radical has not yet been detected nor characterized in the laboratory, and its chemistry is missing from current astrochemical models and databases such as KIDA\footnote{\url{https://kida.astrochem-tools.org/}} and UMIST.\footnote{\url{https://www.umistdatabase.com/}} Consequently, understanding the formation of HNSO first requires elucidating how NSO itself is produced. Combinatorially, three elementary reactions can, in principle, yield NSO on grains:
\begin{align}
  \ce{NO + S &-> NSO} \\
  \ce{SO + N &-> NSO} \label{reac:N_addition} \\
  \ce{NS + O &-> NSO}, \label{reac:O_addition} 
\end{align}
where the reaction \ref{reac:N_addition} was originally suggested in \citet{sanz-novo_discovery_2024}. Elemental two-body processes are assumed for reactions on grains because the ice matrix effectively acts as a third body.\cite{Pantaleone_2020, ferrero_where_2023,molpeceres_reaction_2023} Among these candidates, the route starting from \ce{NO + S} is considered unfavorable, as it would require a significant heavy-atom rearrangement to place sulfur in the central position. \textbf{This assessment is consistent with previous thermochemical studies showing that isomers with an \ce{NSO} structural motif are more stable than those with an \ce{SNO} motif, owing to the ability of the central sulfur atom to effectively expand its valency.\cite{mendez_thermodynamic_2014}} In contrast, the \ce{NS + O} and \ce{SO + N} channels are, in principle, plausible. Both NS and SO are well-established interstellar radicals\cite{Turner1995,Lique2006,laas_modeling_2019,NS_first,1994ApJ...422..621M} also abundant in G+0.693,\cite{sanz-novo_discovery_2024} that can encounter diffusing O and N atoms (of which N is faster due to the lower binding energy on ASW), which have been shown to migrate on ASW at 10 K,\cite{Minissale2013, Minissale2016a,Shimonishi2018, Molpeceres2020, lee2014, Pezzella2018,Zaverkin2021} although more slowly than H. The feasibility of such heavy-atom diffusion arises from the wide distribution of adsorption sites on ASW, a heterogeneity that astrochemical models have only recently begun to represent in a general way.\cite{Grassi2020,furuya_framework_2024}

The interstellar detection of HNSO is therefore timely in several respects. First, it establishes a new benchmark for the chemical complexity of small interstellar molecules, as it contains four different elements and exhibits a relatively high abundance, 6$\times$10$^{-10}$ with respect to \ce{H2}, i.e. only a factor $\sim$ 5 lower than the ubiquitous \ce{SO2} molecule in G+0.693. Second, its formation cannot be explained solely through hydrogenation chemistry and instead requires the inclusion of heavy-atom reactions to account for its synthesis. Third, HNSO serves as an excellent molecule to test the impact of multibinding approaches in astrochemical models, as its reaction network is small enough to be treated in a controlled manner. 

The aim of this work is to understand, from an astrochemical perspective, how HNSO forms, to explain its presence in G+0.693, and to evaluate to what extent its chemistry and detectability may apply to other interstellar environments. The article is structured as follows. Section~\ref{sec:methods} presents the quantum chemical calculations used to determine the surface parameters implemented in our astrochemical models, with the results of such investigation shown in Section \ref{sec:qc}. Section~\ref{sec:model} introduces the modeling framework, describes the resulting chemical abundances, and discusses their astrophysical implications. Finally, Section~\ref{sec:conclusions} summarizes the main findings derived from both the quantum chemical and astrochemical modeling analyses.

\section{Computational Methodology} \label{sec:methods}

\subsection{Reactivity on dust grains}

\begin{figure*}[ht]
  \centering

  \begin{minipage}[b]{0.30\linewidth}
    \centering
    \includegraphics[width=\linewidth]{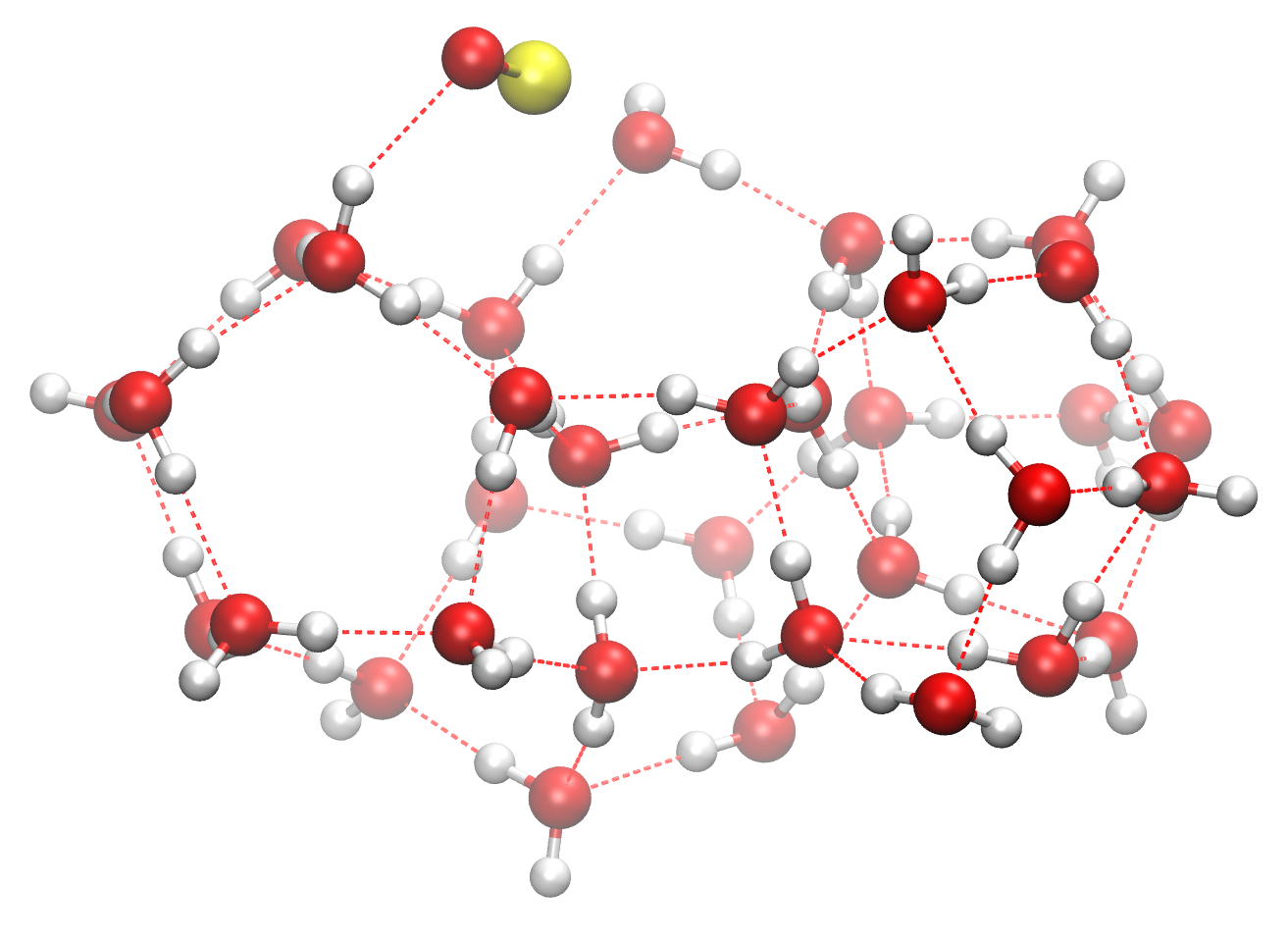}
    \vspace{0.5em}
    \small Pocket site
  \end{minipage}
  \hfill
  \begin{minipage}[b]{0.30\linewidth}
    \centering
    \includegraphics[width=\linewidth]{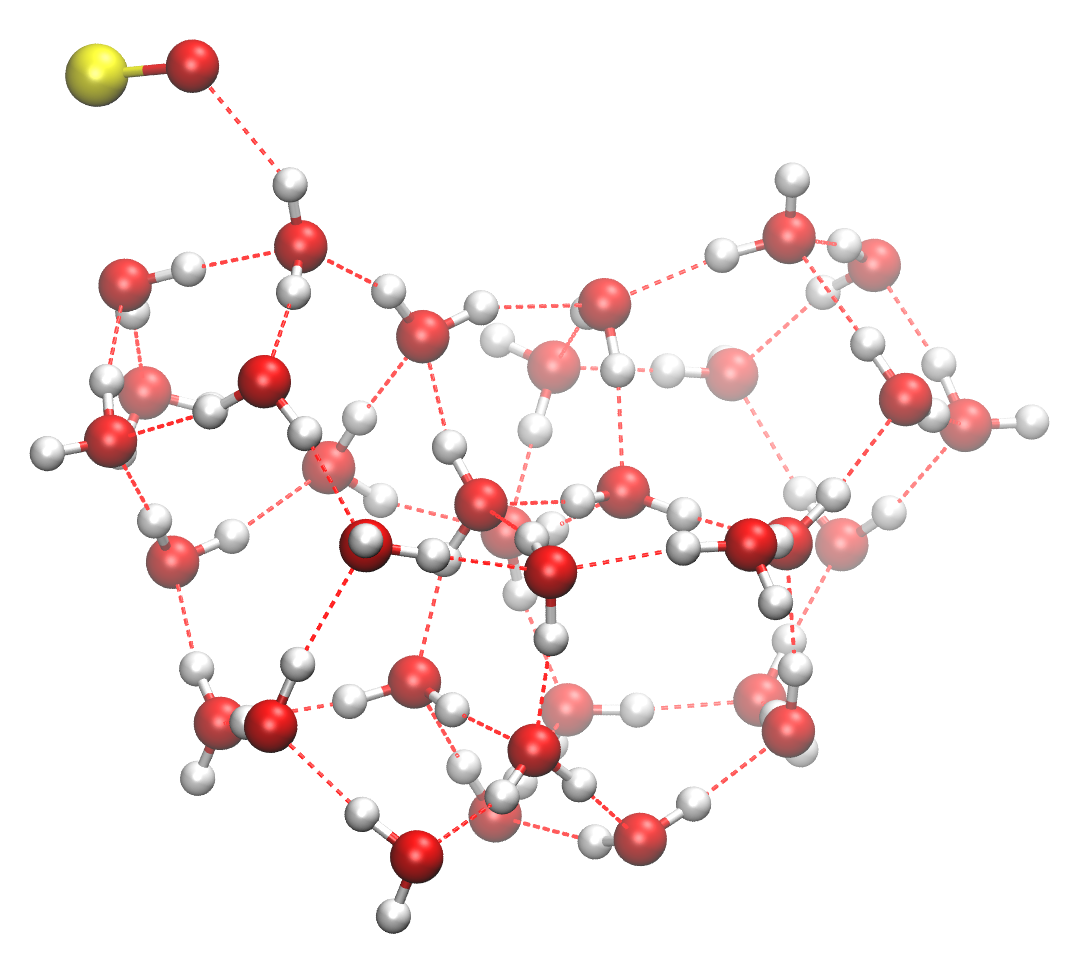}
    \vspace{0.5em}
    \small dH site
  \end{minipage}
  \hfill
  \begin{minipage}[b]{0.30\linewidth}
    \centering
    \includegraphics[width=\linewidth]{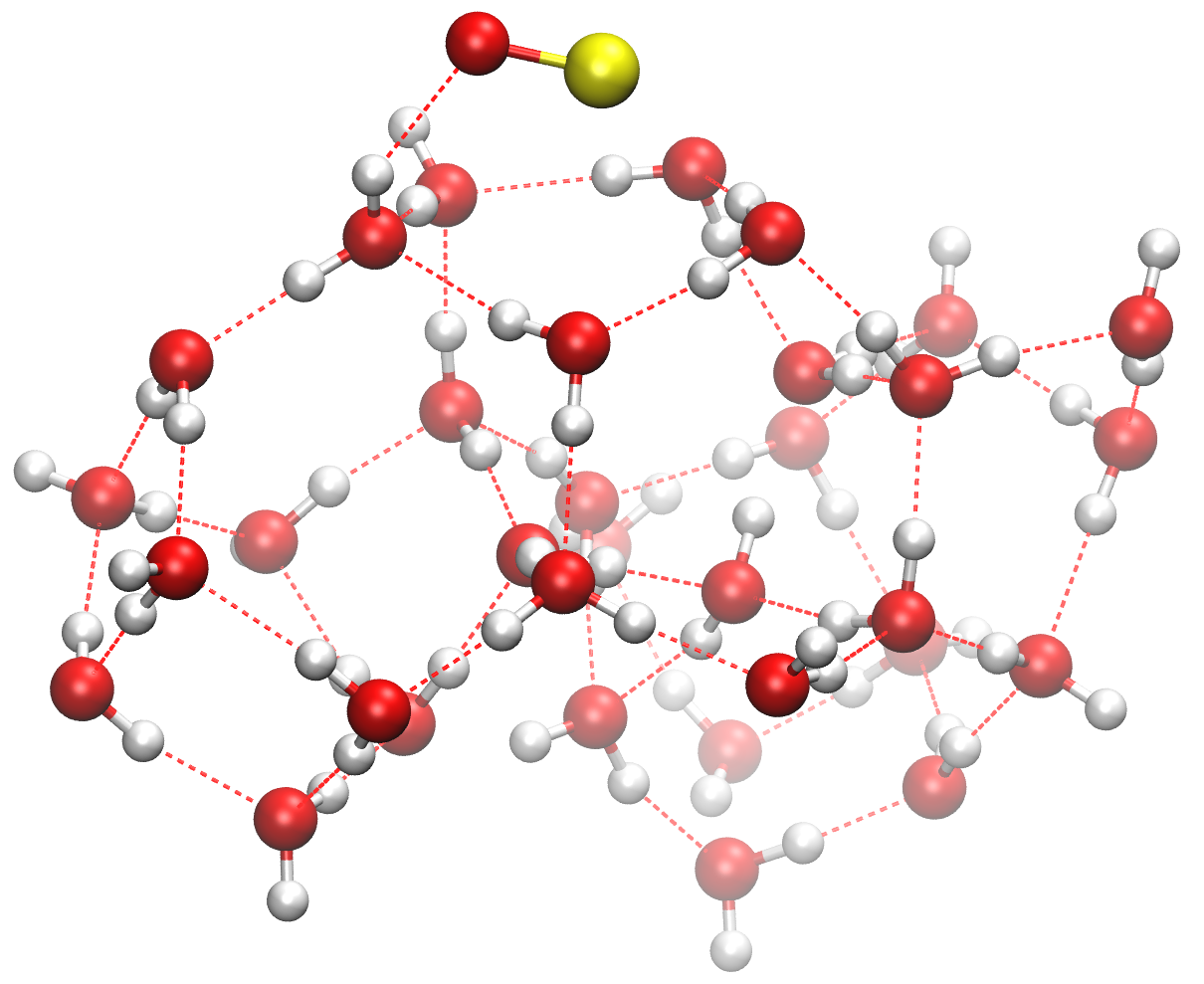}
    \vspace{0.5em}
    \small Pentamer site
  \end{minipage}
  \caption{Examples of the three binding sites on the cluster model of amorphous solid water (ASW) considered in this work. Red dotted lines indicate the hydrogen-bond network. The SO molecule is shown as a representative adsorbate.}
  \label{fig:binding_sites}
\end{figure*}

We performed density functional theory (DFT) calculations to investigate the reactivity leading to the formation of the title molecule. Specifically, we adopted a dual-level approach: geometry optimizations and vibrational frequency (Hessian) calculations were carried out at the $\omega$B97M-D4-gCP/def2-SVP level of theory,\cite{mardirossian_b97x-v_2014, najibi_dft_2020, Caldeweyher2019, Sure2013, Weigend2005} and single-point energy refinements were performed using $\omega$B97M-D4/def2-TZVPPD. \textbf{The latter energy refinement is not gCP corrected, owing to the significantly larger based set used in the computations}. Further improvements through high-level single-determinant wavefunction methods were deemed unfeasible due to the pronounced multiconfigurational character of many of the reactions studied. Likewise, multireference calculations become prohibitively expensive when explicit water molecules are included. Consequently, a broken-symmetry DFT approach\cite{neese_definition_2004} was employed to describe the low-spin reactive channels, providing a practical balance between accuracy and computational cost. \textbf{\citealt{trabelsi} previously characterized the electronic structure of radical \ce{NSO} and \ce{SNO} by means of multireference methods, providing one-dimensional cuts of the potential energy surface. These do not show any appreciable barriers along the reaction coordinate, a behavior that we can duplicate with our broken symmetry DFT approach. } Since most of the reactions investigated are either barrierless or involve low activation barriers, this level of theory is adequate for our purposes. All calculations were performed with the \textsc{Orca} program package, employing resolution-of-identity and seminumerical techniques for efficient evaluation of the Coulomb (RIJ) and exchange (Chain-of-Spheres, COSX) terms.\cite{neese_software_2022, izsak_overlap_2011, neese_improvement_2003, neese_efficient_2009}

We explored the formation pathways of HNSO on amorphous solid water (ASW) using a representative cluster of 33 water molecules, previously employed in related studies.\cite{Rimola2018, Perrero2022, del_valle_formation_2024} The first step in modeling these reactions consisted of placing the SO and NS radicals on three distinct binding sites of the ASW cluster. Each of these sites, labeled \textit{Pocket}, \textit{dH}, and \textit{Pentamer}, each represent a characteristic hydrogen-bonding environment within the ASW structure, as illustrated in Figure~\ref{fig:binding_sites} and its caption. \textbf{From the totally optimized structures}, we derived the binding energies of the species as:

\begin{equation} \label{eq:bindenergies}
  BE = H_{\rm adsorbate} + H_{\rm cluster} - H_{\rm cluster+adsorbate}
\end{equation}
where $H$ represents the enthalpy at zero kelvin, that is, the sum of the electronic and zero-point energies (ZPE). \textbf{The protocol involves placing the radicals on top of the preoptimized water cluster, and relaxing the structure, taking $H_{\rm adsorbate}$ (the remaining term) from a gas phase calculation. } Once the adsorption minima were located, we investigated atom-addition reactions at each site. We began with reactions~\ref{reac:N_addition} and~\ref{reac:O_addition}, deliberately excluding competing processes, which will be discussed in the following sections. Attacks on both atomic centers of each diatomic molecule were considered, resulting in 12 distinct reactions (3 adsorption sites $\times$ 2 reactants, SO and NS, $\times$ 2 atomic addition sites per reactant). \textbf{The binding energies of the distinct products of reactions are obtained similarly, with the optimized structures of the reaction serving as $H_{\rm cluster+adsorbate}$. We must note however, that the calculation arising from three binding sites should not be confused with a rigorous characterization of the binding energy distribution, although our values fall within it. }

The protocol used to study these reactions was the same in all cases. First, starting from the adsorption minima, the N or O atom was placed at an initial distance of approximately 4.0~\AA\ from the adsorbate to mimic co-adsorption of both species. Second, the broken-symmetry wavefunction for the reactant complex was obtained and subsequently used as input to ensure continuity along the relaxed potential energy surface (PES) scans performed for each of the 12 reaction coordinates. The resulting scans were then analyzed to determine whether the reactions were barrierless. For those showing a barrier, the transition state was located through conventional saddle-point optimization. Finally, the endpoints of each scan were fully optimized to yield the reference reactant and product states, and their ZPEs were obtained from Hessian calculations. Reaction and activation energies, when applicable, were then computed as:
\begin{align}
  \Delta H^{\circ,R} &= H_{\rm PS} - H_{\rm RS} \\
  \Delta H^{\circ,A} &= H_{\rm TS} - H_{\rm RS}
\end{align}
\textbf{where PS, RS and TS stand for product state, reactant state and transition state, respectively.}

A second set of reactions was examined starting from the products of the previous ones, as described in Sections~\ref{sec:hydrogenation} and~\ref{sec:other_reactions}. For these reactions, the general protocol remained nearly identical, although specific details are discussed in the corresponding sections. 

\subsection{Quantum chemical treatment of isomerism}

A key feature of HNSO is its conformational flexibility. So far, only the \textit{cis} form has been detected in the ISM,\cite{sanz-novo_discovery_2024} but the \textit{trans} conformer could also be present in space, as observed for other high-energy stereoisomers,\cite{Taquet2017, Agundez2019, sanz-novo_conformational_2025} , yet it lacks available rotatonal data to enable its astronomical search. To assess the possibility of \textit{cis}–\textit{trans} isomerization, we performed exploratory simulations of tunneling-mediated interconversion.\citep{GarciadelaConcepcion2021, GarciadelaConcepcion2022}. For this purpose, we evaluated the feasibility of isomerization in the gas phase by computing the isomerization rate constant, $k^{\rm iso}$, as: 

\begin{equation}
   k^{\rm iso} = \kappa \frac{k_BT}{h} \exp\left(\--\frac{\Delta G^{\neq}}{RT}\right)
   \label{eq:rate_constant}
\end{equation}
The \textit{trans}-\textit{cis} isomerization rate constants and times were recently determined by \citet{jiang_deciphering_2025} and, as shown in Section \ref{sec:isomerization} their values are retained owing to their high precision. In the above expression, and for c/t isomerization, $\Delta G^{\neq}$ represents the free energy barrier, obtained from electronic structure calculations. These were performed at the CCSD(T)-F12/VTZ-F12//$\omega$B97M-D4-gCP/def2-SVP level of theory,\citep{knizia_simplified_2009} while we did not consider isomerization on the surface, as we discussed in Section \ref{sec:isomerization}. The parameter $\kappa$ denotes the tunneling transmission coefficient, which can be evaluated with different levels of sophistication. Since our calculations consistently indicate \textit{cis}–\textit{trans} conversion is intrinsically slow and only weakly dependent on the tunneling model employed, we report rate constants using the computationally efficient asymmetric Eckart approximation, expressed as:

\begin{equation}
\resizebox{\linewidth}{!}{$
\kappa(T) = \frac{1}{k_B T} \exp\left(-\frac{\Delta G^\ddagger}{RT}\right) 
\int_0^\infty P(E) \exp\left(-\frac{E}{RT}\right) \, dE
$}
\label{eq:transmission_coefficient_Eckart}
\end{equation}
where $P(E)$ is the tunneling probability calculated through an assymetric Eckart barrier, whose analytic form can be consulted, for example at \citealt{johnston_tunnelling_1962} or the original reference.\cite{Eckart1930}

\section{Quantum chemical results}  \label{sec:qc}

\subsection{Nitrogen and oxygen addition} 

\begin{table}[h!]
  \caption{Reaction energies and activation energies, when applicable, of the nitrogen and oxygen atom addition reactions studied in this work. All energies are reported in \kcalmol. BL means barrierless. \label{tab:firstT}}
  \centering
  \resizebox{\linewidth}{!}{%
  \begin{tabular}{lccc}
    \hline
    Reaction & Binding site & $\Delta H^{\circ,R}$ & $\Delta H^{\circ,A}$ \\
    \hline
    \ce{SO + N -> NSO} & Pocket & -70.8 & BL \\
                       & dH     & -71.9 & BL \\
                       & Pentamer & -71.3 & BL \\
    \hline
    \ce{SO + N -> NOS} & Pocket$^{\textrm{a}}$ & -37.8$^{\textrm{a}}$ & 14.7 \\
                       & dH     & -16.6 & 16.1 \\
                       & Pentamer & -41.1$^{\textrm{a}}$ & 15.6 \\
    \hline
    \ce{NS + O -> NSO} & Pocket & -80.4 & -0.3$^{\textrm{b}}$ \\
                       & dH     & -82.1 & BL \\
                       & Pentamer & -79.7 & BL \\
    \hline
    \ce{NS + O -> ONS} & Pocket & -78.7 & BL \\
                       & dH     & -78.0 & BL \\
                       & Pentamer & -77.3 & BL \\
    \hline
  \end{tabular}%
  }
  \begin{flushleft}
  \footnotesize
  $^{\textrm{a}}$ The NOS molecule in this binding site dissociates into $^{2}$NO + $^{1}$S, where the $^{1}$S atom chemisorbs on \ce{H2O}.\citep{di_genova_hot_2025} \\
  $^{\textrm{b}}$ A small diffusion barrier at the 2$\zeta$ level submerges after electronic energy correction using the def2-TZVPPD basis set.
  \end{flushleft}
\end{table}

\begin{table}[h!]
  \caption{Binding energies (BE; in \kcalmol and K in parenthesis) of the species involved in the atom addition reactions shown in Section \ref{sec:qc}. \textbf{We note that an average over three binding sites does not preclude a more in-depth investigation of the binding energy distribution of the molecules.}} \label{tab:be_first}
  \centering
  \begin{tabular}{lcc}
    \hline
    Species & Binding site & BE \\
    \hline
    SO & Pocket & 5.5 (2755) \\
       & dH     & 2.7 (1368) \\
       & Pentamer & 5.1 (2755) \\
    \hline
    \multicolumn{2}{c}{\textbf{Average SO:}} & 4.4 (2222) $\pm$ 1.5 \\
    \hline    
    NS & Pocket & 6.0 (3007) \\
       & dH     & 4.1 (2038) \\
       & Pentamer & 6.2 (3108) \\
    \hline
    \multicolumn{2}{c}{\textbf{Average NS:}} & 5.4 (2717) $\pm$ 1.2 \\
    \hline    
    NSO & Pocket   & 6.6 (3330) \\
        & dH       & 4.8 (2438) \\
        & Pentamer & 7.3 (3682) \\
    \hline
    \multicolumn{2}{c}{\textbf{Average NSO:}} & 6.3 (3150) $\pm$ 1.3 \\
    \hline
    NOS$^{\textrm{a}}$ & dH & 1.9 (971) \\
    \hline    
    ONS & Pocket   & 4.7 (2364) \\
        & dH       & 2.1 (1039) \\
        & Pentamer & 3.4 (1727) \\
    \hline
    \multicolumn{2}{c}{\textbf{Average ONS:}} & 3.4 (1710) $\pm$ 1.3 \\
    \hline
  \end{tabular}
  \begin{flushleft}
  \footnotesize
  $^{\textrm{a}}$ Only binding site for NOS.
  \end{flushleft}
\end{table}

To explore potential grain-surface pathways for the formation of HNSO, we first identified small interstellar species that are both abundant and capable of forming an –S– moiety. Within these constraints, SO and NS emerge as the most plausible candidates, as described in Section~\ref{sec:intro}. The adsorption of SO and NS on ASW results in physisorbed species with average binding energies of 4.4~$\pm$~1.5~\kcalmol\ (2222~K) for SO and 5.4~$\pm$~1.2~\kcalmol\ (2717~K) for NS (see Table~\ref{tab:be_first}). The reported uncertainties correspond to the standard deviation of the individual values. These results indicate that the diffusion of SO and NS is unlikely to compete with that of atomic N and O, \textbf{which have binding energies of approximately 720 or 400~K depending on the literature, \cite{Minissale2016a,Molpeceres2020} and 1300~K,\cite{Minissale2016a} respectively.} This supports the conclusion that N and O diffusion are the main processes driving reactions~\ref{reac:N_addition} and~\ref{reac:O_addition}.

\subsubsection{The $^{4}$N + $^{3}$SO route}

We begin our exploration of the HNSO reaction network by examining the addition of an N atom to SO. The SO molecule, in its electronic ground state (\(^{3}\Sigma\)), was placed on the three adsorption binding sites, and the two possible attack directions of a (\(^{4}S\))-N atom were analyzed in the doublet channel, assuming antiparallel spins between SO and N. Reactions proceeding through the high-spin (sextet) configuration, which could lead to abstraction channels such as \ce{SO + N -> NS + O} or \ce{SO + N -> NO + S}, must be endothermic because they produce electronically excited fragments. The quartet state, although potentially reactive, shows pronounced multiconfigurational character, making DFT inadequate for its reliable description. As this study focuses on low-spin channels that promote radical coupling, the investigation of the quartet potential energy surface (PES) is deferred to future work. 

The doublet state was generated using a broken-symmetry electronic wavefunction: first, the high-spin state was converged, and then the low-spin wavefunction was obtained by explicitly enforcing a spin configuration corresponding to separated spins on SO and N.\citep{neese_definition_2004} The same computational protocol was applied to Reaction~\ref{reac:O_addition}. The results for the two possible additions, namely \textbf{Reaction \ref{reac:N_addition} and}:
\begin{align}
  \ce{SO + N &-> NOS} \label{reac:SO-N_NOS}
\end{align}
are shown in the first two entries of Table \ref{tab:firstT}

The results for both reactions are summarized in Table~\ref{tab:firstT}, together with the binding energies (BE) of all sulfur-bearing species involved, listed in Table~\ref{tab:be_first}. Reaction~\ref{reac:N_addition} is consistently exothermic and barrierless across all binding sites, making it highly favorable. \textbf{In contrast, Reaction~\ref{reac:SO-N_NOS} exhibits significant activation barriers at all sites,} with little variation among them. Notably, the formation of \ce{NOS} shows some dependence of $\Delta H^{\circ,R}$ on the binding site. This arises because \ce{NOS} is unstable on certain sites and spontaneously dissociates into \ce{NO + S}, with the sulfur atom in the singlet state chemisorbing onto water.\citep{di_genova_hot_2025} However, in the dH site, \ce{NOS} remains stable. Despite this, the large activation barrier and the lack of quantum tunneling render the nitrogen-addition to the oxygen atom pathway highly improbable and ultimately irrelevant under astrophysical conditions. Consequently, the formation of \ce{NOS} will not be considered in subsequent discussions. 

Given the barrierless and exothermic nature of Reaction~\ref{reac:N_addition}, we conclude that it is the \textbf{most favorable pathway within the present network} for the \ce{SO + N} reaction and model it with a branching ratio (\(\alpha\)) of 1.0. Finally, although not further considered in this work, the BE of the \ce{NOS} radical in the sites where it is stable is found to be 1.9~\kcalmol\ (Table~\ref{tab:be_first}).

\subsubsection{The $^{3}$O + $^{2}$NS route} \label{sec:o-ns}

The second possible route to form the parent radical \ce{NSO} is through the addition of an O atom to \ce{NS}. As in Reaction~\ref{reac:N_addition}, two possible attack directions were considered, \textbf{both Reaction \ref{reac:O_addition} and}:
\begin{align}
  \ce{NS + O &-> ONS}
  \end{align}
Both reactions are exothermic, with comparable $\Delta H^{\circ,R}$ values. Moreover, both channels are barrierless, in contrast with the results presented in the previous subsection, indicating that their occurrence depends mainly on the relative orientation of the \ce{NS} molecule and the incoming O atom, which can be considered random on an anisotropic potential such as that of ASW surfaces. In this work, we focus on the chemistry leading to HNSO formation and do not examine in detail the reaction network stemming from the \ce{ONS} radical. A preliminary analysis of the spin density of \ce{ONS} suggests that further radical–radical H-addition reactions could yield undetected interstellar species such as \ce{NSOH} and \ce{N(SH)O}, owing to the delocalized nature of the unpaired spin. This behavior contrasts with that of \ce{NSO}, where the spin population is largely localized on the nitrogen atom, favoring the predominant formation of HNSO (Section~\ref{sec:hydrogenation}). However, a more detailed investigation would be necessary before reaching firm conclusions.

For the modeling of Reaction~\ref{reac:O_addition} (Section \ref{sec:model}), we adopt a branching ratio of $\alpha = 0.5$ to represent the competition between the \ce{NS + O -> NSO} and \ce{NS + O -> ONS} channels. It should be noted that this treatment corrects only for the potential overproduction of \ce{NSO} from the \ce{NS + O} reaction, but does not account for the sulfur sink in \ce{NOS} and related species, which slightly increases the sulfur available for other compounds. Finally, as in the case of the \ce{NOS} radical, the \ce{ONS} radical—although not considered in subsequent analyses—has a binding energy of 3.4~\kcalmol.

\subsection{Hydrogenation of NSO} \label{sec:hydrogenation}

The next step in the formation of HNSO is the addition of a hydrogen atom to the \ce{NSO} radical. The evolution of the \ce{NOS} and \ce{ONS} radicals is not considered as already mentioned. Other channels are included, when relevant, only indirectly, as they influence the formation efficiency of the target molecule in the astrochemical models. The \ce{NSO + H} reaction can proceed through five competing pathways:

\begin{align}
  \ce{NSO + H &-> cis-HNSO} \\
  \ce{NSO + H &-> trans-HNSO} \\
  \ce{NSO + H &-> cis-NSOH} \\
  \ce{NSO + H &-> trans-NSOH} \\
  \ce{NSO + H &-> N(SH)O} 
\end{align}
In the above reactions, competition among the different channels makes it difficult to isolate individual processes, as both barrierless and activated pathways coexist within a relatively narrow orientational space (given that \ce{NSO} is a small radical). To investigate these reactions, we followed a strategy similar to that described by \citet{ferrero_formation_2023}. In their approach, barrierless pathways are identified by monitoring the spontaneous conversion of a pre-reactive complex into products during a force minimization, starting from a configuration in which the reactants are initially placed at a large separation distance.

\begin{table*}[h!]
  \caption{Energetic parameters for all \ce{NSO + H} reactions studied in this work, as a function of the binding site. The “Moiety” column specifies the atom targeted by the incoming H atom, while the “Orientation” column indicates the corresponding stereochemical configuration. All energies are given in~\kcalmol. \label{tab:hydrogenation}}
  \centering
  \begin{tabular}{ccccccc}
    \hline
    Binding site & Moiety & Orientation & Barrierless reaction? & Product & $\Delta H^{\circ,R}$ & $\Delta H^{\circ,A}$ \\
    \hline
    Pocket & N & Cis   & \ding{51} & cis-HNSO    & -104.8 & BL \\
    Pocket & N & Trans & \ding{55} & NP        & N/A  & N/A \\
    Pocket & O & Cis   & \ding{51} & cis-HNSO    & -104.8$^{a}$ & BL \\
    Pocket & O & Trans & \ding{55} & trans-NSOH    & -84.6 & 3.8 \\
    Pocket & S & S     & \ding{55} & N(SH)O    & -56.4 & 0.8 \\
    dH     & N & Cis   & \ding{51} & cis-HNSO    & -102.9 & BL \\
    dH     & N & Trans & \ding{51} & trans-HNSO    & -100.9 & BL \\
    dH     & O & Cis   & \ding{51} & cis-HNSO    & -102.9$^{a}$ & BL \\
    dH     & O & Trans   & \ding{55} & cis-NSOH  & -93.9 & 4.4 \\
    dH     & S & S   & \ding{55} & N(SH)O  & -50.6 & 0.6 \\
    Pentamer & N & Cis & \ding{51} & cis-HNSO & -102.7 & BL \\
    Pentamer & N & Trans & \ding{51} & trans-HNSO & -103.5 & BL \\
    Pentamer & O & Cis & \ding{51} & cis-HNSO & -102.7$^{a}$ & BL \\
    Pentamer & O & Trans & \ding{51} & cis-HNSO & -102.7$^{a}$ & BL \\
    Pentamer & S & S & \ding{55} & trans-NSOH  & -87.2 & 0.9$^{b}$ \\
    \hline
  \end{tabular}

  \begin{flushleft}
  \footnotesize
  $^{a}$-The reactant and product states are assumed to be essentially equivalent to the reaction on the N-cis moiety. $^{b}$-The product experiences an intermolecular proton transfer mediated isomerization reaction with the water matrix, \ce{N(SH)O -> trans-NSOH} NP: No product.
  \end{flushleft}
\end{table*}

\begin{figure}
    \includegraphics[width=\linewidth]{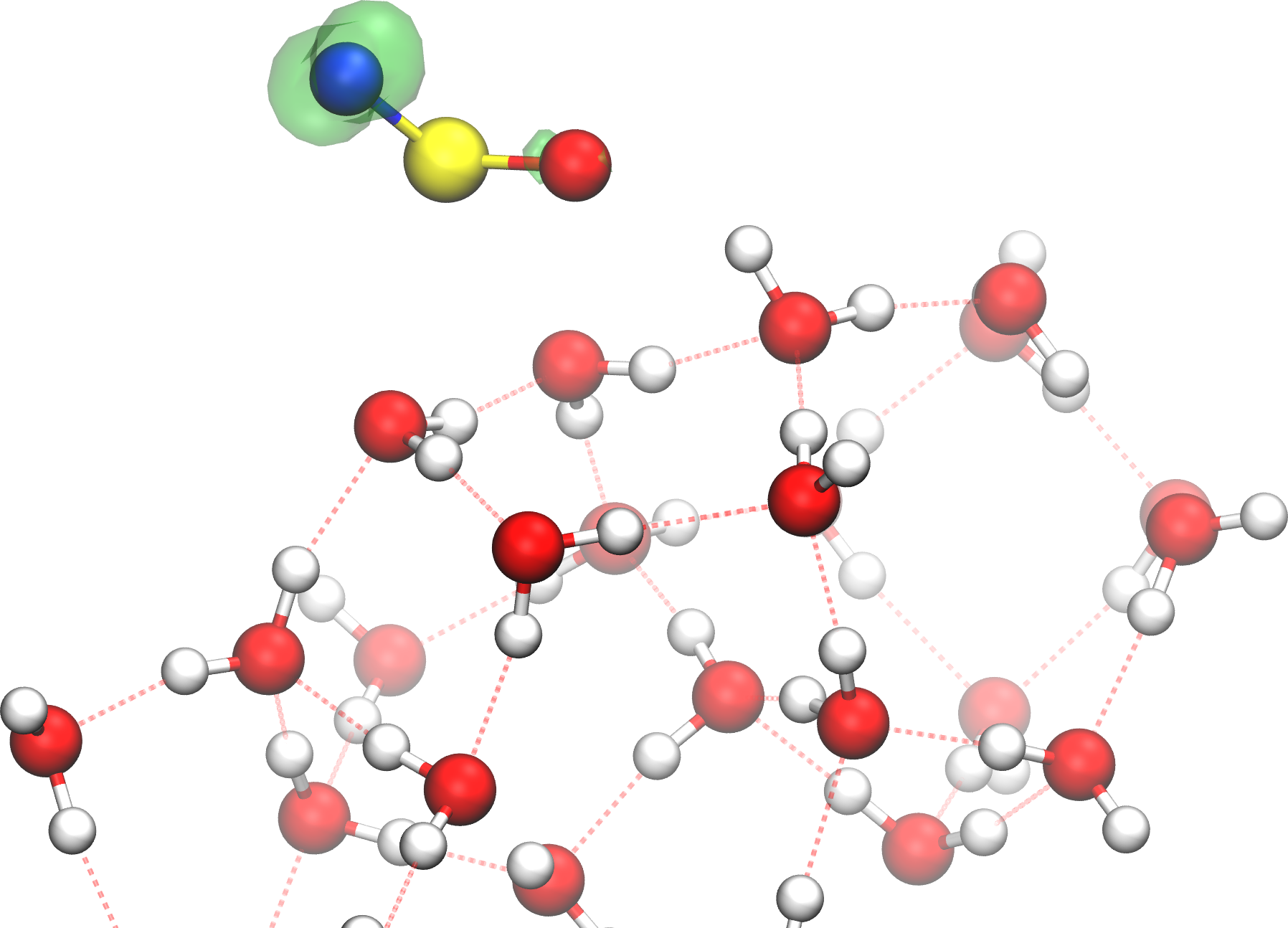}
  \caption{Spin density of the NSO radical obtained with an isovalue of 0.02 a.u.}
  \label{fig:spin_density}
\end{figure}

We start the protocol placing H atoms in various relative orientations with respect to the \ce{NSO} radical and at different binding sites. Additionally, we slightly extended the protocol of \citet{ferrero_formation_2023} by performing a constrained geometry optimization of the pre-reactant complex, fixing the distance between the H atom and the nearest atom of the \ce{NSO} molecule at 3.0~Å. After this constrained step, the H atom was released, and a full geometry optimization was carried out to determine whether the reaction proceeded without an activation barrier. The results of this analysis are summarized in Table~\ref{tab:hydrogenation}. 

Overall, we find that \textit{c}-\ce{HNSO}, the isomer detected in the ISM, is consistently formed across all binding sites and over a wide range of initial orientations. For instance, \textit{c}-\ce{HNSO} is produced when the H atom approaches the N end of the \ce{NSO} molecule in the \textit{cis} conformation, but also when approaching the O end in the same orientation, or even in less favorable configurations such as the O–\textit{trans} end at the pentamer site. This clearly indicates that \textit{c}-\ce{HNSO} is the most favored isomer, a preference that can be rationalized in terms of both electronic and geometric effects. As shown in Figure~\ref{fig:spin_density}, the spin density of the \ce{NSO} radical at the pentamer site is mainly localized at the N end. Combined with the bent, V-shaped geometry of the molecule, this facilitates the formation of \textit{c}-\ce{HNSO} even when the initial approach occurs near the O atom. 

The second product that forms without activation barriers on the ASW surface is \textit{t}-\ce{HNSO}, which arises when the H atom approaches the N end of the \ce{NSO} molecule in the \textit{trans} configuration. However, in our limited sampling, there are cases where \textit{c}-\ce{HNSO} is still preferred over \textit{t}-\ce{HNSO}. For example, at the Pocket site, no accessible orientation leads to \textit{t}-\ce{HNSO} formation, making \textit{c}-\ce{HNSO} the only viable product. Attacks at the O end generally involve sizable activation barriers and are therefore expected to contribute less to the overall product distribution, similar to attacks at the S site. Interestingly, the barriers for the latter are considerably lower, suggesting that the formation of \ce{N(SH)O}, while rare, cannot be entirely ruled out. Nevertheless, to maintain the chemical model presented in Section~\ref{sec:model} tractable and to avoid introducing additional species with poorly characterized chemistry, these alternative products are not explicitly included in the reaction network. Finally, we note an interesting case involving hydrogenation at the S site of the pentamer configuration, which leads to the formation of \textit{t}-\ce{NSOH} instead of \ce{N(SH)O}. This unconventional outcome results from a favorable hydrogen-bond arrangement that facilitates a proton relay following initial \ce{N(SH)O} formation. We expect this mechanism could occur more frequently when considering additional binding sites, offering an alternative route to produce \textit{t}-\ce{NSOH}. However, as discussed above, and for the purposes of this work, \textit{c}-\ce{HNSO} remains the clearly preferred product of the \ce{NSO + H} reaction.

\subsection{Other reactions} \label{sec:other_reactions}

In order to complete the reaction network surrounding NSO and HNSO we have investigated a number of additional reactions to be later included in our rate equation models. In particular, we considered reactions of NSO with other atoms (N,O), and H-abstraction reactions from HNSO.

In the first place, we studied the \ce{NSO + O} reaction in the low-spin channel following the same method that we used for the previous hydrogenation in the Pocket site, not finding any appreciable reactivity. On the contrary, the reaction with N atoms on the same site spontaneously (barrierless) forms \ce{N2} through the following reaction:

\begin{equation}
  \ce{NSO + N -> SO + N2} \label{eq:n2}
\end{equation}

with a $\Delta H^{\circ,R}$ of $-$155.4 \kcalmol, i.e., an enormous reaction energy, following the formation of the very stable \ce{N2} molecule. This route acts a promising competing route to hydrogenation, reducing the abundance of HNSO.

Finally, we complete the trifecta of studied reactions with the H-abstraction in HNSO:
\begin{align}
    \ce{cis-HNSO + H &-> NSO + H2} \\
    \ce{trans-HNSO + H &-> NSO + H2} \label{eq:h-abs-t}
\end{align}
Interestingly, this reaction is sensitive to the isomeric form of \ce{HNSO}, with gas-phase reaction energies of 1.3~\kcalmol\ (endothermic) and –1.5~\kcalmol\ \textbf{exothermic} when starting from \ce{cis-HNSO} and \ce{trans-HNSO}, respectively. We did not further investigate the endothermic channel. For Reaction~\ref{eq:h-abs-t}, its study on the model ice yields an activation enthalpy of $\Delta H^{\circ,A} \sim 13$~\kcalmol, corresponding to a very high barrier that precludes its viability. \textbf{While quantum tunneling might be invoked as a driver for the reactions, the barriers are probably too high to be competitive with H diffusion away of the reaction center. Moreover, given that the almost null exothermicty of the reaction, one could also invoke the back reactions as viable, as they would be equally affected by tunneling, ultimately making H-abstraction of HNSO an irrelevant chemical process in the HNSO network.} Therefore, among the additional reactions considered in this section, only Reaction~\ref{eq:n2} is relevant to the chemistry of \ce{HNSO} and is the one included in our astrochemical models.

\textbf{Alternative formation routes of \ce{HNSO} by means of direct combination reaction such as \ce{NH + SO -> HNSO}, \ce{SH + NO -> HNSO} and \ce{NS + OH -> HNSO} are not considered in our study. \citep{Hassani} studied these pathways by means of different single- and multi--reference computational methods. Reactions \ce{SH + NO -> HNSO} and \ce{NS + OH -> HNSO} involve the barrierless formation of \ce{HSNO} and \ce{HOSN}, respectively. In both cases, the hydrogen atom must subsequently rearrange through a series of intermediate steps to yield the desired product \ce{HNSO} (either the cis or trans isomer). These intermediate steps involve a series of energy barriers that, considering the icy grain as a third body, will not be overcome. The remaining alternative (\ce{NH + SO}), though reported exothermic and barrierless, would necessarily involve the diffusion of NH on the grain surfaces, whose higher diffusion energy (e.g. \citealt{molpeceres_enhanced_2024}) makes this route less competitive, at least on \ce{H2O} ice and under cold, thermalized, conditions.} 

\subsection{cis/trans-HNSO isomerization} \label{sec:isomerization}

\begin{figure}
    \includegraphics[width=\linewidth]{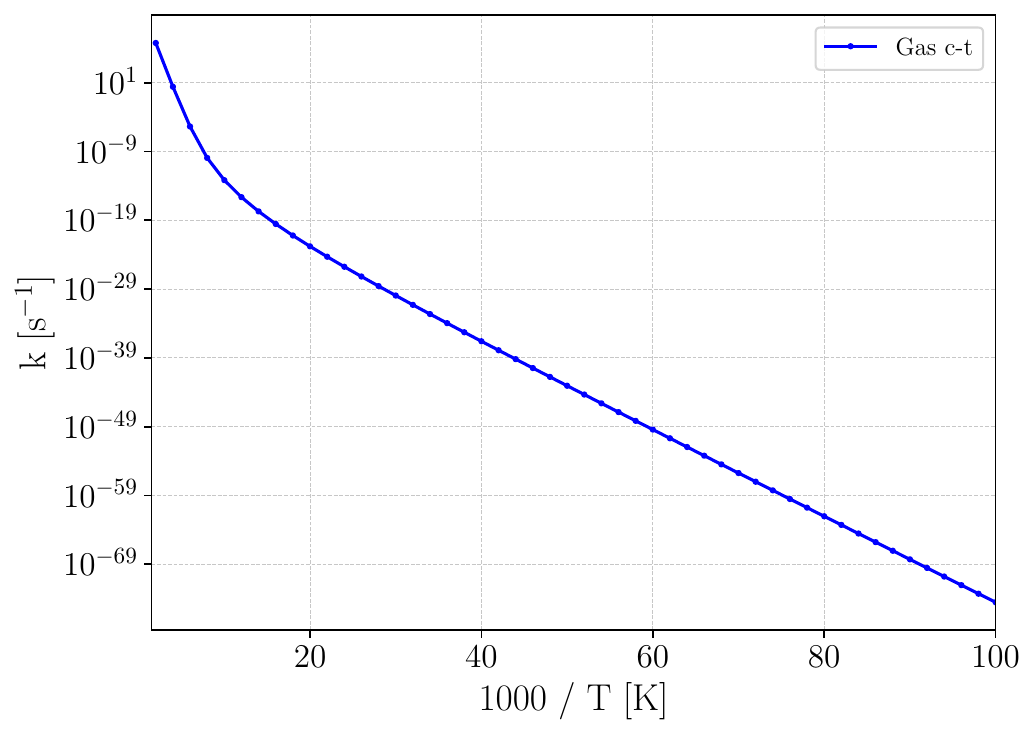}
  \caption{Eckart corrected isomerization rate constants for the gas phase \textit{cis}-\text{trans} isomerization.}
  \label{fig:isomer}
\end{figure}
We conclude our quantum chemical investigation of HNSO reactivity by examining its possible isomerization, both in the gas phase and on the ASW surface. For the latter, we focus on the dH binding site, where both isomers can form. In the gas phase, the energy difference between the two isomers is 2.8~\kcalmol\ (1413~K), with \textit{cis}-\ce{HNSO} being the most stable form. This separation is smaller than typical values reported for other interstellar imines.\citep{GarciadelaConcepcion2021} The isomerization barriers between the \textit{cis} and \textit{trans} forms are, however, very high: 14.5~\kcalmol\ (7305~K) for the \textit{cis}~$\rightarrow$~\textit{trans} process and 11.7~\kcalmol\ (5892~K) for the reverse. On the ASW surface, these barriers remain similar, at 12.9~\kcalmol\ (6495~K) and 10.9~\kcalmol\ (5502~K), respectively. 

These computed energetics are consistent with those reported by \citet{jiang_deciphering_2025}, who found \textit{trans}–\textit{cis} barriers of approximately 10~\kcalmol. Based on their experimental results, we derive a \ce{trans-HNSO -> cis-HNSO} rate constant of $6.92\times10^{-4}$~s$^{-1}$ at 3~K, corresponding to a half-life of about 16~minutes.\cite{jiang_deciphering_2025} This extremely short lifetime challenges the presence of \textit{t}-\ce{HNSO} under interstellar conditions provided that \textit{cis}-\textit{trans} isomerization is slow as demonstrated in the next paragraphs. 

Previous studies of isomerization reactions on ice surfaces\cite{Molpeceres2021b,molpeceres_hydrogenation_2025} have shown that direct isomerization is often hindered by hydrogen bonding, which restricts molecular motion and can raise the effective activation barrier by several orders of magnitude despite a marginal decrease in the activation energies. It is unclear whether this effect also applies to \textit{t}-\ce{HNSO} or if spontaneous isomerization could occur on grains. In this work, we assume that surface isomerization does not take place, since reproducing the calculations of \citet{jiang_deciphering_2025} on ices would be computationally prohibitive. Nevertheless, this assumption does not affect our conclusions regarding the detectability of \textit{t}-\ce{HNSO}, as its very short gas-phase half-life at near-zero temperature already makes its detection challenging. Therefore,  for astrochemical modeling, the \textit{cis}~$\rightarrow$~\textit{trans} isomerization channel is also considered. We computed its rate constants using an Eckart tunneling model as indicated in Section \ref{sec:methods}. The results, shown in Figure~\ref{fig:isomer}, confirm that the process is intrinsically slow. Therefore, in our astrochemical model, we adopt the experimental rate constant from \citet{jiang_deciphering_2025} in the 10–90~K temperature range and the instanton rate constant reported in their Supplementary Information above 90~K for the \ce{trans-HNSO -> cis-HNSO} reaction, and fit our own derived rate constants for the the \ce{cis-HNSO -> trans-HNSO} reaction constant to a modified Arrhenius expression (see Section~\ref{sec:model_desc}).

\section{Chemical model and astrophysical implications} \label{sec:model}

\subsection{Description of the chemical model} \label{sec:model_desc}

To evaluate the astrochemical relevance of the quantum chemical results presented in the previous sections, we performed exploratory simulations of HNSO formation under different interstellar conditions. The specific models employed are described later in this section. For this purpose, we used the three-phase gas–grain code \textsc{Rokko},\cite{furuya_water_2015} coupled to the latest version of our modified chemical network,\cite{Molpeceres2025c} which is greatly based on the three-phase reaction network developed by \citet{garrod_three-phase_2013}. A list of the modifications and additions to the \citet{garrod_three-phase_2013} network on which our work builds can be found in Appendix~A of \citet{Molpeceres2025c}; here, we briefly summarize only the features specific to this study. \textbf{As we will show during this section, our model does not fully include a full sulfur chemistry network pertaining G+0.693, nor it includes several of the important physical processes characteristic of this cloud. Therefore, the agreement between the models and observations can be regarded as qualitative, with significant uncertainties still needed to be resolved in the gas-gas and gas-grain interplay of HNSO and more generally, sulfur chemistry in G+0.693.}

A distinctive aspect of this work is the incorporation of the full set of reactions relevant to HNSO, including those discussed above as well as standard gas-phase and grain-surface processes. The details of these new reactions and their implementation are provided in Section~\ref{sec:input_model}. We reiterate here that our models are essentially exploratory and the rate constants for the gas-phase processes are included for completeness of the reaction network and might significantly vary from the actual values. Nevertheless, the chemistry of HNSO in our models is dominated by grain-surface processes, which were constrained through the quantum chemical calculations presented in Section~\ref{sec:qc}.

\begin{table}[h!]
  \caption{Elemental abundances (with respect to hydrogen nuclei) at the beginning of each chemical model. The notation A(B) represents A$\times$10$^{\rm B}$.} \label{tab:initial_abundances}
  \centering
  \begin{tabular}{lc}
    \hline
    Element & Fractional abundance \\
    \hline
    \ce{H2} & 0.5    \\
    H       & 1(-5)  \\
    He      & 9(-2)  \\
    N       & 6.2(-5) \\
    O       & 2.4(-4) \\
    \ce{C+} & 1.7(-4) \\
    S$^{+}$ & 1.5(-5)  \\
    Si$^{+}$ & 1.8(-6) \\
    Fe$^{+}$ & 1.0(-8) \\
    Na$^{+}$ & 2.3(-7) \\
    Mg$^{+}$ & 2.3(-6) \\
    Cl$^{+}$ & 1.0(-9) \\
    P$^{+}$  & 7.8(-8) \\
    \hline
  \end{tabular}
\end{table}

The three models introduced below share the same initial abundances. We adopt as starting abundances the high-metallicity values proposed by \citet{wakelam_efficiency_2021}, with all elements except hydrogen assumed to be in their atomic form, either neutral or ionized depending on their ionization potentials (see Table~\ref{tab:initial_abundances}). Hydrogen is considered predominantly molecular (\ce{H2}), with a small atomic fraction ($1\times10^{-5}$) included to represent the steady-state abundance of H atoms produced by cosmic-ray–induced \ce{H2} dissociation.\cite{Goldsmith2005} 

Sulfur is assumed to be less depleted on dust grains at the start of the simulations, consistent with previous observations toward galactic center clouds and recent theoretical modeling on S-bearing chemistry toward G+0.693.\cite{martin-hernandez_iso_2002,rodriguez-fernandez_iso_2005,sanz-novo_discovery_2024} A visual extinction of 10~mag is imposed in all cases, representing a cloud well shielded from the interstellar radiation field. The models incorporate several non-thermal desorption mechanisms, namely cosmic-ray–induced heating, photodesorption (with a yield of $1\times10^{-3}$), and chemical desorption. For the latter, a constant efficiency of 1\% per reactive event is used.

\begin{table}[h!]
  \caption{Differing conditions in the three astrochemical models studied in this work. Gas density (n$_{H}$) is set to 2$\times$10$^{4}$ cm$^{3}$ in all models. The notation A(B) represents A$\times$10$^{\rm B}$. } \label{tab:diff_models}
  \centering
  \resizebox{\linewidth}{!}{\begin{tabular}{lcccc}
    \hline
    Label & T$_{\rm gas}$ (K) & T$_{\rm dust}$ (K) & $\zeta$ (s$^{-1}$) & Multibinding? \\
    \hline
    Model A & 10 & 10 & 1.3(-17) &  \ding{55}   \\
    Model B & 10 & 10 & 1.3(-17) &  \ding{51}   \\
    Model C & 100 & 15 & 1.3(-16) &  \ding{51}   \\
    \hline
  \end{tabular}
  }
\end{table}

In addition to the common initial conditions described above and elsewhere,\cite{molpeceres_hydrogenation_2025} we varied several physical parameters to progressively introduce additional complexity into the models. Three chemical models were run, designed to explore the influence of a multibinding treatment of the binding energies\cite{furuya_framework_2024} and of a set of more energetic physical conditions as elevated cosmic-ray ionization rates and temperatures. The differing parameters among the three models are summarized in Table~\ref{tab:diff_models}.  Model~A represents the simplest case, employing a single binding energy per species and cold gas and dust temperatures. Model~B introduces a multibinding treatment of the binding energies while maintaining the same cold temperatures as Model~A. Finally, Model~C features a higher cosmic-ray ionization rate (1.3$\times$10$^{-16}$ s$^{-1}$), a warmer gas temperature (100~K), and increased dust temperature (15~K). The physical environment in G+0.693 where HNSO is found is surely more energetic, but the here presented models cannot reproduce these conditions. For example, we tested higher $\zeta$ in our exploratory single binding model, finding a significant decrease of all our considered species and an overall very poor agreement with observations. We attribute this to the inability of our model to fully account for the CMZ particularities, like shock induced desorption, radiolysis, or mantle chemistry. Therefore, an increase of one order of magnitude in $\zeta$ is sufficient to evaluate the impact of this parameter on molecular abundances, postponing (and strongly advocating for) a more detailed modeling of G+0.693 to future works.

The inclusion of a multibinding treatment in Models~B and~C accounts for the variability in binding energies arising from different adsorption sites on ASW surfaces, as discussed in Section~\ref{sec:qc}. In this work, the multibinding approach was applied to O, N, and S atoms, using the standard binding energies from \citet{garrod_three-phase_2013} (1320, 720, and 2600~K, respectively) and assuming a distribution with a full width at half maximum (FWHM) of 0.2 relative to the mean value. We also tested initial conditions with reduced average binding energies for the N atom \textbf{approximatedly 400 K},\cite{Molpeceres2020, Shimonishi2018} but found no significant effect on the resulting abundances, as the multibinding scheme naturally encompasses these variations at low temperatures. All other species in the reaction network are treated using a single binding energy. Overall, the multibinding approach provides a more realistic representation of grain-surface chemistry compared to the single binding energy assumption used in Model~A.

\subsection{Input parameters used in the chemical model} \label{sec:input_model}

\begin{table*}
  \caption{Reactions added to the modified \citet{garrod_three-phase_2013} reaction network.\cite{Molpeceres2025c}\label{tab:additions}. In addition to these reactions, other processes like adsorption, desorption, cosmic-ray induced desorption\cite{hasegawa_three-phase_1993} are part of the reaction network. The notation A(B) represents A$\times$10$^{\rm B}$. }
  \centering
  \resizebox{\linewidth}{!}{\begin{tabular}{lcccc}
    \hline
    Reaction & Type  & $\alpha$ & $\beta$ & $\gamma$ \\
    \hline
    \ce{SO(i) + N(i) -> NSO(i)} & Two Body Surface &  1.00 & 0.00 & 0.00 \\
    \ce{NS(i) + O(i) -> NSO(i)} & Two Body Surface & 0.50 & 0.00 & 0.00 \\
    \ce{NS(g) + OH(g) -> NSO(g) + H(g)} & Neutral-neutral gas & 1.00(-11) & 0.00 & 0.00 \\
    \ce{NSO(i) + H(i) -> cis-HNSO(i)} & Two Body Surface & 0.66$^{a}$ & 0.00 & 0.00 \\
    \ce{NSO(i) + H(i) -> trans-HNSO(i)} & Two Body Surface & 0.33$^{a}$ & 0.00 & 0.00 \\
    \ce{NSO(i) + N(i) -> SO(i) + N2(i)} & Two Body Surface & 1.00 & 0.00 & 0.00 \\
    \ce{SO(g) + NH2(g) -> cis-HNSO(g) + H(g)} & Neutral-neutral gas & 5.00(-12) & 0.00 & 0.00 \\
    \ce{SO(g) + NH2(g) -> trans-HNSO(g) + H(g)} & Neutral-neutral gas & 5.00(-12) & 0.00 & 0.00 \\
    \ce{NSO(g) + H(g) -> HNO(g) + S(g)} & Neutral-neutral gas & 1.00(-11) & 0.00 & 0.00 \\
    \ce{NSO(g) + N(g) -> SO(g) + N2(g)} & Neutral-neutral gas & 1.00(-11) & 0.00 & 0.00 \\
    \ce{NSO(g) + C$^{+}$(g) -> NSO$^{+}$(g) + C(g)} & Ion-Molecule gas & 1.00(-9) & 0.00 & 0.00 \\
    \ce{NSO(g) + He$^{+}$(g) -> NSO$^{+}$(g) + He(g)} & Ion-Molecule gas & 1.00(-9) & 0.00 & 0.00 \\
    \ce{HNSO$^{+}$(g) + NH3(g) -> NSO(g) + NH4$^{+}$(g)} & Ion-Molecule gas & 1.00(-9) & 0.00 & 0.00 \\
    \ce{NSO$^{+}$(g) + H2 -> HNSO$^{+}$(g) + H(g)} & Ion-Molecule gas &  1.00(-9) & 0.00 & 0.00 \\
    \ce{NSO(g) + HCO$^{+}$(g) -> HNSO$^{+}$(g) + CO(g)} & Ion-Molecule gas & 1.00(-9) & 0.00 & 0.00 \\
    \ce{NSO(g) + H3$^{+}$(g) -> HNSO$^{+}$(g) + H2(g)} & Ion-Molecule gas & 1.00(-9) & 0.00 & 0.00 \\
    \ce{NSO(g) + N2H$^{+}$(g) -> HNSO$^{+}$(g) + N2(g)} & Ion-Molecule gas & 1.00(-9) & 0.00 & 0.00 \\
    \ce{HNSO$^{+}$(g) + H2(g) -> H2NSO$^{+}$(g) + H(g)} & Ion-Molecule gas & 1.00(-9) & 0.00 & 0.00 \\
    \ce{HNSO$^{+}$(g) + e$^{-}$ -> NSO(g) + H(g)} & Dissociative recombination & 1.00(-7) & 0.00 & 0.00 \\
    \ce{cis-HNSO(g) +  C$^{+}$(g) -> HNSO$^{+}$(g) + C(g)} & Ion-Molecule gas$^{b}$ & 1.80(-9) & -3.68(-1) & -3.30(-1) \\
    \ce{trans-HNSO(g) +  C$^{+}$(g) -> HNSO$^{+}$(g) + C(g)} & Ion-Molecule gas & 4.97(-9) & -4.60(-1) & -9.90(-2) \\
    \ce{cis-HNSO(g) + He$^{+}$(g) -> NSO$^{+}$(g) + H(g) + He(g)} & Ion-Molecule gas & 2.95(-9) & -3.69(-1) & -3.30(-1) \\
    \ce{trans-HNSO(g) + He$^{+}$(g) -> NSO$^{+}$(g) + H(g) + He(g)} & Ion-Molecule gas & 8.11(-9) & -4.60(-1) & -9.90(-2) \\
    \ce{cis-HNSO(g) + HCO$^{+}$(g) -> H2NSO$^{+}$(g) + CO(g) } & Ion-Molecule gas & 1.29(-9) & -3.69(-1) & -3.30(-1) \\
    \ce{trans-HNSO(g) + HCO$^{+}$(g) -> H2NSO$^{+}$(g) + CO(g) } & Ion-Molecule gas & 3.54(-9) & -4.60(-1) & -9.90(-2) \\
    \ce{cis-HNSO(g) + N2H$^{+}$(g) -> H2NSO$^{+}$(g) + N2(g) } & Ion-Molecule gas & 1.29(-9) & -3.69(-1) & -3.30(-1) \\
    \ce{trans-HNSO(g) + N2H$^{+}$(g) -> H2NSO$^{+}$(g) + N2(g) } & Ion-Molecule gas & 3.54(-9) & -4.60(-1) & -9.90(-2) \\
    \ce{cis-HNSO(g) + H3$^{+}$(g) -> H2NSO$^{+}$(g) + H2(g) } & Ion-Molecule gas & 3.39(-9) & -3.69(-1) & -3.30(-1) \\
    \ce{trans-HNSO(g) + H3$^{+}$(g) -> H2NSO$^{+}$(g) + H2(g) }& Ion-molecule gas & 9.30(-9) & -4.60(-1) & -9.90(-2) \\
    \ce{H2NSO$^{+}$(g) + e$^{-}$ -> cis-HNSO(g) + H(g)}  & Dissociative recombination & 3.33(-8) & 0.00 & 0.00 \\
    \ce{H2NSO$^{+}$(g) + e$^{-}$ -> trans-HNSO(g) + H(g)} & Dissociative recombination & 3.33(-8) & 0.00 & 0.00 \\
    \ce{H2NSO$^{+}$(g) + e$^{-}$ -> NSO(g) + H2(g)}  & Dissociative recombination & 3.33(-8) & 0.00 & 0.00 \\
    \ce{H2NSO$^{+}$(g) + NH3(g) -> cis-HNSO(g) + NH4$^{+}$(g)}  &Ion-Molecule gas & 5.00(-10) & 0.00 & 0.00 \\
    \ce{H2NSO$^{+}$(g) + NH3(g) -> trans-HNSO(g) + NH4$^{+}$(g)}  & Ion-Molecule gas & 5.00(-10) & 0.00 & 0.00 \\
    \ce{NSO + h$\nu$ -> N + SO}$^{c}$ &
    Photodissociation interstellar field & 2.13(-9) & 0.00 & 2.00 \\
    \ce{NSO(g) + h$\nu$ -> NS(g) + O(g)} &
    Photodissociation interstellar field & 2.13(-9) & 0.00 & 2.00 \\
    \ce{NSO(g) + h$\nu$ -> N(g) + SO(g)} &
    Photodissociation cosmic-ray photon & 2.77(3) & 0.00 & 0.00 \\
    \ce{NSO(g) + h$\nu$ -> NS(g) + O(g)} &
    Photodissociation cosmic-ray photon & 2.77(3) & 0.00 & 0.00 \\
    \ce{NSO$^{+}$(g) + h$\nu$ -> N(g) + SO$^{+}$(g)} &
    Photodissociation interstellar field & 3.43(-10) & 0.00 & 2.00 \\
    \ce{NSO$^{+}$(g) + h$\nu$ -> O(g) + NS$^{+}$(g)} &
    Photodissociation interstellar field & 3.43(-10) & 0.00 & 2.00 \\
    \ce{NSO$^{+}$(g) + h$\nu$ -> N(g) + SO$^{+}$(g)} &
    Photodissociation cosmic-ray photon & 2.30(2) & 0.00 & 0.00 \\
    \ce{NSO$^{+}$(g) + h$\nu$ -> O(g) + NS$^{+}$(g)} &
    Photodissociation cosmic-ray photon & 2.30(2) & 0.00 & 0.00 \\
    \ce{cis/trans-HNSO(g) + h$\nu$ -> NSO(g) + H(g)} &
    Photodissociation interstellar field & 1.42(-9) & 0.00 & 2.00 \\
    \ce{cis/trans-HNSO(g) + h$\nu$ -> NSO(g) + H(g)} &
    Photodissociation cosmic-ray photon & 1.85(3) & 0.00 & 0.00 \\
    \ce{cis/trans-HNSO(g) + h$\nu$ -> SO(g) + NH(g)} &
    Photodissociation interstellar field & 1.42(-9) & 0.00 & 2.00 \\
    \ce{cis/trans-HNSO(g) + h$\nu$ -> SO(g) + NH(g)} &
    Photodissociation cosmic-ray photon  & 1.85(3) & 0.00 & 0.00 \\
    \ce{cis/trans-HNSO(g) + h$\nu$ -> NS(g) + OH(g)} &
    Photodissociation interstellar field & 1.42(-9) & 0.00 & 2.00 \\
    \ce{cis/trans-HNSO(g) + h$\nu$ -> NS(g) + OH(g)} &
    Photodissociation cosmic-ray photon  & 1.85(3) & 0.00 & 0.00 \\
    \ce{HNSO$^{+}$(g) + h$\nu$ -> H(g) + NSO$^{+}$(g)} &
    Photodissociation interstellar field & 6.86(-10) & 0.00 & 2.00 \\
    \ce{HNSO$^{+}$(g) + h$\nu$ -> H(g) + NSO$^{+}$(g)} &
    Photodissociation cosmic-ray photon & 4.60(2) & 0.00 & 0.00 \\
    \ce{H2NSO$^{+}$(g) + h$\nu$ -> H2(g) + NSO$^{+}$(g)} &
    Photodissociation interstellar field & 3.43(-10) & 0.00 & 0.00 \\
    \ce{H2NSO$^{+}$(g) + h$\nu$ -> H2(g) + NSO$^{+}$(g)} &
    Photodissociation cosmic-ray photon & 2.30(2) & 0.00 & 2.00 \\
    \ce{H2NSO$^{+}$(g) + h$\nu$ -> H(g) + HNSO$^{+}$(g)} &
    Photodissociation interstellar field & 3.43(-10) & 0.00 & 2.00 \\
    \ce{H2NSO$^{+}$(g) + h$\nu$ -> H(g) + HNSO$^{+}$(g)} &
    Photodissociation cosmic-ray photon & 2.30(2) & 0.00 & 0.00 \\
    \ce{cis-HNSO(g) -> trans-HNSO(g)} &
    Unimolecular isomerization & 1.12(-1) & 0.00 & 1.69(3) \\
    \ce{trans-HNSO(g) -> cis-HNSO(g)} &
    Unimolecular isomerization & 6.92(-4)/2.26(-3)$^{d}$ & 0.00 & 0.00 \\
    \hline
  \end{tabular}}
  \begin{flushleft}
\footnotesize
$^{a}$~Calculated from the ratios obtained in Table~\ref{tab:hydrogenation} (see text).  
$^{b}$~The ratio between the isomeric ion–molecule destruction rates is based on the RDP (see text).  
$^{c}$~Photodissociation rates are assumed to be those of SO for neutral species and \ce{SH+} for cationic species, taken from the Leiden database,\cite{heays_photodissociation_2017} and weighted by the number of considered channels. A shielding function with $\gamma = 2.0$ is applied to all photodissociation reactions.  
$^{d}$~Rate constants taken from \citet{jiang_deciphering_2025} at 3~K and 90~K. The latter corresponds to instanton calculations reported in their Supplementary Information.
For all the gas-phase reactions not mentioned in the previous notes we use rate constants of $1\times10^{-11}$~cm$^{3}$~s$^{-1}$ for neutral–neutral reactions, $1\times10^{-9}$~cm$^{3}$~s$^{-1}$ for ion–molecule reactions, and $1\times10^{-7}$~cm$^{3}$~s$^{-1}$ for dissociative recombination, divided by the number of postulated channels.
  \end{flushleft}
\end{table*}

The most significant additions to the chemical models used in this work are the reactions involving \ce{NSO} and \ce{HNSO}, particularly the heavy-atom addition processes described in Section~\ref{sec:qc}. A summary of the newly included reactions is provided in Table~\ref{tab:additions}. As noted in the caption, the table does not list all the physical processes implemented in the model but focuses primarily on the newly incorporated chemical reactions.

\subsubsection{Gas phase reactions: Ion-molecule, unimolecular conversion and photorates}

\begin{table}[h!]
  \caption{Dipole moments ($\mu$, in Debye) and diagonal components of the polarizability tensor ($\alpha_{ii}$, in \AA$^{3}$) for cis-HNSO and trans-HNSO}  \label{tab:dipole}
  \centering
  \begin{tabular}{lcccc}
    \hline
    Species & $\mu$ & $\alpha_{xx}$ & $\alpha_{yy}$ & $\alpha_{zz}$ \\
    \hline
    cis-HNSO & 0.91 & 1.41 & 1.75 & 2.60  \\
    trans-HNSO & 3.50 & 1.41 & 1.71 & 2.66  \\
    \hline
  \end{tabular}
\end{table}

Although our quantum chemical investigation focuses exclusively on grain-surface chemistry, we also incorporated gas-phase processes into our chemical models, albeit with a lower level of detail and using effective rate theories with assumed products and branching ratios. To construct the gas-phase reaction network, we distinguished five classes of reactions: neutral–neutral, ion–molecule, dissociative electron recombination, photodissociation (driven by both the interstellar radiation field and cosmic-ray–induced photons), and unimolecular isomerization. 

For the first three categories, and as a general rule, rate constants were taken as the collisional rate divided by the number of available reaction channels. We adopted collisional rate constants of $1\times10^{-11}$~cm$^{3}$~s$^{-1}$ for neutral–neutral reactions, $1\times10^{-9}$~cm$^{3}$~s$^{-1}$ for ion–molecule reactions, and $1\times10^{-7}$~cm$^{3}$~s$^{-1}$ for dissociative recombination. The different rates obey to conventional choices in the astrochemical literature and the differences in orders of magnitude between the different processes stem from the different physicochemical forces driving the reactive collision, e.g., dipole-induced dipole, ion-dipole/induced dipole or ion-electron recombination.  An exception to this general treatment is the ion–molecule reactivity of the \textit{cis}- and \textit{trans}-\ce{HNSO} stereoisomers, as it is well established that the relative abundances of stereoisomers are influenced by their differing destruction reactivities, a concept known as the Relative Dipole Principle (RDP).\cite{Shingledecker2020} For these processes we explicitly computed the Su–Chesnavich capture rate.\cite{su_parametrization_1982}

\begin{align}
  k_\mathrm{D,x<2} &= k_\mathrm{L}(0.4767x + 0.6200) \label{eq:su1} \\
  k_\mathrm{D, x>2} &= k_\mathrm{L}\left[ \frac{(x + 0.5090)^2}{10.526} + 0.9754 \right] \label{eq:su2}
\end{align}
where:
\begin{align}
  x &= \frac{\mu_\mathrm{D}}{\sqrt{2\alpha k_\mathrm{B}T}}, \\
  k_\mathrm{L} &= 2\pi e \sqrt{\frac{\alpha}{\mu}},
\end{align}
Here, $\mu_\mathrm{D}$ denotes the total dipole moment of the neutral reactant, $\alpha$ its scalar polarizability, $\mu$ the reduced mass of the reactants, and $e$ the elementary charge. The dipole moments and diagonal components of the polarizability tensors for \textit{c}-\ce{HNSO} and \textit{t}-\ce{HNSO} were obtained at the $\omega$B97M-D4/def2-TZVPPD level of theory, and the corresponding values are listed in Table~\ref{tab:dipole}. For ease of implementation in every astrochemical model, instead of directly using Equations~\ref{eq:su1} and~\ref{eq:su2}, we fitted the rate constants derived from these expressions to a modified Arrhenius equation; the resulting parameters are given in Table~\ref{tab:additions}. 

In the absence of specific photochemical data for NSO derivatives, we adopt photodissociation rates by analogy: SO rates for neutral species and \ce{SH+} rates for cations, scaled by the number of available photodissociation channels in each case. Rates are taken from the Leiden database (\url{https://home.strw.leidenuniv.nl/~ewine/photo/})\cite{heays_photodissociation_2017}. This uniform treatment removes any possibility of preferential photodestruction among isomers and will certainly overestimate or underestimate absolute photodissociation rates. Nevertheless, when assessing the total destruction rates for all species included in the additional reactions considered here, we find that photodissociation is several orders of magnitude less relevant than ion–molecule pathways under the shielded conditions adopted (A$_{V}$=10 mag). Consequently, uncertainties in the assumed photorates have a negligible impact on our results. 

Finally, we included the rate constants obtained in Section \ref{sec:isomerization} for the isomerization of cis-HNSO and trans-HNSO. In the model, we only included isomerization in the gas phase, as the effect of an ice matrix reduces the rate constants as discussed in Section \ref{sec:isomerization} and the chemical timescale of species is dominated by hydrogenation rather than by isomerization. For reference a H atom accretes on average at a rate of one atom per day at canonical $\zeta$ (1.3$\times$10$^{-17}$ s$^{-1}$).\cite{Wakelam2017} We note that \citet{jiang_deciphering_2025} only report rate constants up to 90 K, and therefore for the inclusion on the models, we recommend to use the rate constants at 90 K for temperatures above this value. Caution is advised when determining cis-HNSO/trans-HNSO ratios at high temperatures, where the rate constants should be extrapolated above the validity range of our fit and \textit{trans}-\textit{cis} rate constants are not available.

\subsubsection{Surface reactions}

We constructed an \ce{NSO}/\ce{HNSO} reaction network based on the results presented in Section~\ref{sec:qc}. In brief, we added two-body surface reactions following the Langmuir–Hinshelwood formalism described by \citet{hasegawa_three-phase_1993} and \citet{ruaud_gas_2016}, assuming reaction–diffusion competition for processes with activation barriers. \textbf{Though the Eley-Rideal gas-grain formalism could also be considered, as all the here considered reactions would be viable, the low coverage of \ce{NO} and \ce{NS} on the ice would make the total reaction rate constant very small.} Our quantum chemical calculations indicate that the hydrogenation of \ce{SO} and \ce{NS} proceeds through distinct numbers of reactive channels. Based on our sampling, we assume that the reaction between \ce{SO} and \ce{N} forms \ce{NSO} with 100\% efficiency, whereas the reaction between \ce{NS} and \ce{O} produces \ce{NSO} and \ce{ONS} in a 2:1 ratio. The \ce{ONS} radical is not explicitly included in the model, as its chemistry remains largely unexplored; its presence is instead implicitly accounted for by the reduced efficiency of the \ce{NS + O -> NSO} reaction.  Additionally, we included the destruction of \ce{NSO} through Reaction~\ref{eq:n2}. The hydrogenation of \ce{NSO} leading to \ce{HNSO} was also implemented, with a 2:1 branching ratio for the formation of \textit{c}-\ce{HNSO} and \textit{t}-\ce{HNSO}, respectively, based on the results summarized in Table~\ref{tab:hydrogenation}. No further reactions involving \ce{HNSO}, such as hydrogen abstraction, whose was deemed uninmportant in Section~\ref{sec:other_reactions}  or additional hydrogenation to form more complex species like \ce{H2NSO}, were included. Nevertheless, their investigation constitutes a promising direction for future studies.

\subsubsection{Average binding energies}

The binding energies of sulfur-bearing species on ASW have been the subject of increasing attention in recent years, both for common S-bearing molecules \citep{perrero_binding_2022} and for sulfur allotropes \citep{perrero_binding_2024}. Among the species considered in this work, and for the chemical models discussed in Section~\ref{sec:model},we determined the binding energies that were still missing from our dataset, namely those of \textit{c}-\ce{HNSO} and \textit{t}-\ce{HNSO}, in addition to the values presented in Table~\ref{tab:be_second}. \textbf{In this section, we consider important to reiterate that our reported binding energies come only from an average over three binding sites, without explicit knowledge of the site population on ASW. Therefore, from the limited sampling we cannot extract reliable binding energy distributions where dedicated works are needed.\cite{tinacci_theoretical_2022, bovolenta_co_2025, bariosco_gaseous_2025, tinacci_theoretical_2023} }

Although the binding energies of both \textit{c}- and \textit{t}-\ce{HNSO} are sufficiently high to make thermal desorption unlikely under typical dense-cloud conditions, it is worth noting the close similarity between the two isomers. This behavior is consistent with our previous findings for formic acid \citep{molpeceres_hydrogen_2022}, where the binding energy distributions of the \textit{trans} and \textit{cis} isomers were found to be essentially equivalent in terms of their adsorption strength.

\begin{table}[h!]
  \caption{Binding energies (BE, in \kcalmol and K in parenthesis) of species different than the ones in Table \ref{tab:be_first} of the new species included in the modified \citet{garrod_three-phase_2013} reaction network.\cite{Molpeceres2025c}. \textbf{We note that an average over three binding sites does not preclude a more in-depth investigation of the binding energy distribution of the molecules.}}  \label{tab:be_second}. 
  \centering
  \begin{tabular}{lcc}
    \hline
    Species & Binding site & BE \\
    \hline
    cis-HNSO & Pocket   & 8.2 (4121) \\
        & dH       & 4.6 (2326) \\
        & Pentamer & 6.6 (3313) \\
    \hline
    \multicolumn{2}{c}{\textbf{Average cis-HNSO:}} & 6.5 (3253) $\pm$ 1.8  \\
    \hline
        trans-HNSO & dH   & 5.4 (2703) \\
        & Pentamer       &  \textbf{10.4} (5220) \\
    \hline
    \multicolumn{2}{c}{\textbf{Average trans-HNSO:}} & 7.9 (3962) $\pm$ 3.5  \\
    \hline
  \end{tabular}
\end{table}

\subsection{Chemical Model Results. Astrophysical Implications}

\begin{figure*}
    \includegraphics[width=\linewidth]{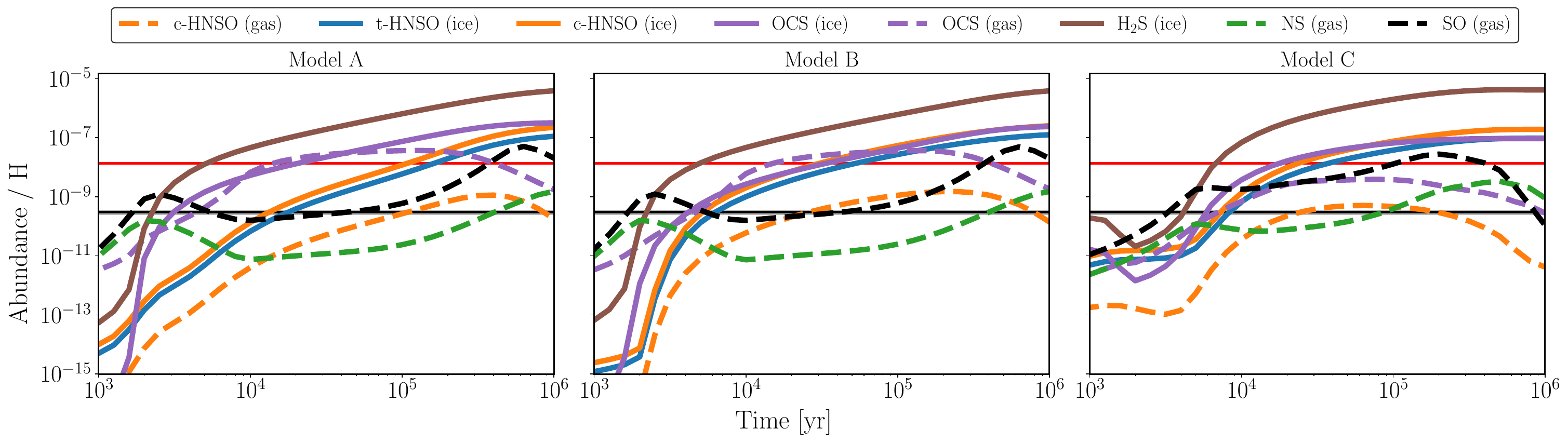}
  \caption{Astrochemical model results for Models A--C see Table \ref{tab:diff_models}. In all models, the abundance of gas trans-HNSO is below 10$^{-19}$ due to spontaneous tunneling conversion.\cite{jiang_deciphering_2025}. The black horizontal bar represents the observed abundance of HNSO  while the red one represents the gaseous OCS abundance, both in in G+0.693.\cite{sanz-novo_discovery_2024}. We note that a good agreement with observations of G+0.693 does not imply that the physical description of the cloud is accurate, see text.}
  \label{fig:isomer_model}
\end{figure*}

The results of the astrochemical models discussed in the previous section are presented in Figure \ref{fig:isomer_model}, highlighting several key points. One of the most surprising aspects of the interstellar detection of HNSO by \citet{sanz-novo_discovery_2024} is the relatively high abundance of such a complex molecule (complex in terms of chemical composition rather than size), despite its small dipole moment of $\sim$0.90~D, as in cis-HNSO. This abundance suggests either highly efficient formation routes, routes that start from very abundant precursors, or a combination of both. These conditions are, in principle, difficult to reconcile with a molecular formula containing H, N, S, and O. Our quantum chemical calculations reveal that efficient grain-surface chemistry can arise from the simplest precursors, namely atomic oxygen and nitrogen. We also tested sulfur-atom diffusion as a potential driver of chemistry, considering that the NS or SO  production could be enhanced by S-atom diffusion, but its contribution was found to be lower than that of O and N, owing to its reduced mobility. The models shown in Figure~\ref{fig:isomer_model} clearly indicate that the increase in HNSO abundance occurs after the freeze-out of SO and NS, around $10^{3}$~years, when the main routes leading to the formation of solid SO and NS originate from the gas phase. This behavior resembles that of icy \ce{H2S} (brown line), which in our model forms primarily through the direct hydrogenation of atomic sulfur at all times. \ce{H2S} is the most abundant sulfur-bearing molecule on ices in our simulations, in good agreement with observational constraints,\cite{Boogert2015} with modeled abundances of 0.5--2.0~\% relative to water in the $10^{4}$--$10^{5}$~yr range. In contrast, OCS formation is largely independent of SO and NS, as it proceeds mainly through the gas-phase reaction \ce{HCS + O -> OCS + H} (where \ce{HCS} originates from \ce{S + CH2 -> HCS + H}) and the grain-surface reaction \ce{S + CO -> OCS} at all times. The dual gas-phase and surface chemistry allows OCS, one of the only icy sulfur-bearing molecules unambiguously identified in space to date\cite{palumbo_solid_1997}, to form via routes that do not compete with those of HNSO, ultimately leading to comparable abundances between the two species. This is likely the most important conclusion of the present work. On ices, our models predict an HNSO abundance comparable to that of OCS, positioning HNSO as a \emph{primary} sulfur reservoir under typical molecular cloud conditions. Gas-phase abundances of OCS and HNSO (both isomers), on the other hand, are not comparable, which is expected since gas-phase formation of OCS remains one of the dominant channels. These results strongly motivate further searches for HNSO in other astronomical environments, as this molecule may represent a key player in the sulfur chemistry of the interstellar medium.

We would like to emphasize the differences arising from the distinct approaches employed in our models. Although the general conclusions discussed above hold for all three models at steady or near–steady state (around 10$^{6}$~years), both ice and gas abundances are significantly affected at earlier times, particularly between 10$^{4}$ and 10$^{5}$~years when the multibinding scheme is applied. This treatment directly influences the gas-phase abundance of HNSO, bringing the modeled values into good agreement with the observational constraints (horizontal line in Figure~\ref{fig:isomer_model}), especially in Model~C. This model, which represents the actual astrochemical history of the molecule a bit better, allows HNSO to reach the same ice abundance as OCS at early times, thereby enhancing its subsequent release into the gas phase. It is important to reiterate, however, that the non-thermal desorption mechanisms responsible for releasing HNSO into the gas phase remain poorly constrained, limiting our ability to accurately predict its detectability in conventional dark clouds. Nevertheless, our model, despite its significant simplifications, is able to reproduce the formation of HNSO in Model~C, and it encourages extending the chemistry leading to HNSO to other interstellar molecules that could form through O or N diffusion on grains, in addition to the more typical hydrogenation pathways. Finally, we tested models incorporating a higher sulfur depletion factor, but these did not reproduce the observed abundances satisfactorily either, reinforcing the idea that sulfur is not severely depleted in energetic environments.\cite{sanz-novo_discovery_2024} From a methodological perspective, our results demonstrate the effectiveness of the multibinding approach for modeling grain-surface chemistry in astrochemical simulations, consistent with the benchmark study of \citet{furuya_framework_2024}, where a similar enhancement of ice \ce{CO2} abundances was reported.

Finally, we observe that in all models, both at low and high gas temperatures, the \textit{trans}–\textit{cis} isomerization makes the abundance of \textit{t}-HNSO negligibly small. This is due to the fast spontaneous interconversion between conformers due to quantum tunneling,\cite{jiang_deciphering_2025} suggesting that detecting this species will be extremely challenging unless additional isomerization mechanisms operate in the region. For comparison, it is known that photon-driven isomerization can bring the abundance of cis-HCOOH to parity with that of the low-energy isomer trans-HCOOH in warm photodissociation regions \citep{cuadrado_trans-cis_2016}, where the high-energy form would otherwise be only residual.\cite{Molpeceres2025c} By analogy, if the low-lying electronic states of \textit{c}-HNSO are favorable for photoisomerization to \textit{t}-HNSO, the latter might be detectable in photodissociation regions, making it a potential tracer of the local photon field. Although this remains speculative, deriving photoisomerization cross sections represents a plausible line of future work, particularly given that the proposed formation mechanism under Earth conditions of \textit{t}-HNSO involves UV irradiation at 254 nm \citep{jiang_deciphering_2025} of \textit{c}-HNSO. In the solid phase, our models predict that the abundance of \textit{t}-HNSO is comparable to that of \textit{c}-HNSO. The reliability of this later prediction, however, depends on two key approximations in our modeling. First, mantle chemistry is not included, meaning that any molecule buried by subsequent ice growth is effectively protected from further reactions. Second, unimolecular isomerizations are neglected in the ice, as molecular motions responsible for isomerism are assumed to be hindered in the solid phase. These two simplifications may overestimate the amount of \textit{t}-HNSO on grains. Nonetheless, they do not affect the conclusions of this work, which rest on the gas-phase abundance of \textit{c}-HNSO and the total HNSO abundance in the ice mantle.

\section{Conclusion} \label{sec:conclusions}

In summary, our results \textbf{suggest} that the formation of HNSO \textbf{can follow} a grain-surface pathway. The combination of quantum chemical and astrochemical modeling allows us to extract several key conclusions, summarized below:

\begin{itemize}
  \item Additions of atomic oxygen and nitrogen to the interstellar radicals \ce{NS} and \ce{SO} efficiently produce the HNSO precursor, \ce{NSO}. Our calculations also show that the positional isomer \ce{ONS} can form from the \ce{NS + O} reaction with a comparable branching ratio to Reaction~\ref{reac:O_addition}. In contrast, formation of the metastable \ce{NOS} radical involves prohibitively high activation barriers and is therefore unlikely.

  \item Hydrogenation of \ce{NSO} readily yields HNSO, with a clear preference for the \textit{cis} (cis-HNSO) conformer on ice surfaces. The \textit{trans} form (trans-HNSO) can also be produced, together with small amounts of radicals such as \ce{N(SH)O} and \ce{NSOH}. Once formed, HNSO ,regardless of the conformer, is resistant to hydrogen abstraction.

  \item We evaluated the spontaneous isomerization of trans-HNSO to cis-HNSO based on the available literature.\cite{jiang_deciphering_2025} Our findings confirm that trans-HNSO is a metastable species under an astronomical prism, that rapidly converts to the more stable cis-HNSO via quantum tunneling. The reverse process (\ce{cis-HNSO -> trans-HNSO}) is extremely slow, making the \ce{cis-HNSO <=> trans-HNSO} equilibrium almost completely displaced to cis-HNSO at low temperatures, and implying that astronomical detection of trans-HNSO will be very difficult.

  \item Our exploratory astrochemical models, which explicitly include the chemistry of \ce{NSO} and \ce{HNSO}, predict that the abundance of HNSO in ices is comparable to that of OCS, a sulfur-bearing molecule unambiguously detected in interstellar ices. This result suggests that HNSO could be a major sulfur reservoir in the solid phase. In the gas phase, however, the presence of efficient gas-phase routes for OCS competing with its destruction and the lack of those for HNSO results in HNSO abundances much lower than those of OCS.

  \item Incorporating a multibinding treatment in our astrochemical models substantially modifies the time evolution of both ice and gas-phase HNSO, increasing its abundance at earlier evolutionary stages of dark clouds and improving agreement with observations. This effect is expected to extend to other molecules formed via oxygen or nitrogen diffusion, as shown by \citet{furuya_framework_2024} for icy \ce{CO2}.
\end{itemize}

Future work should explore the broader chemical landscape accessible from the HNSO molecular formula. This includes not only successive hydrogen additions but also the potential formation of high-energy isomers along the hydrogenation sequence, involving reactions of H atoms with transient radicals such as \ce{N(SH)O} or \ce{NSOH}, which are not yet incorporated into current chemical networks. From an astrochemical standpoint, targeted searches for HNSO in other star-forming regions would be particularly valuable, given the relatively high abundances predicted by our models. In addition, a significantly improved characterization of the energetic chemistry of HNSO in G+0.693, where it is ultimately detected.

%%%%%%%%%%%%%%%%%%%%%%%%%%%%%%%%%%%%%%%%%%%%%%%%%%%%%%%%%%%%%%%%%%%%%
%% The "Acknowledgement" section can be given in all manuscript
%% classes.  This should be given within the "acknowledgement"
%% environment, which will make the correct section or running title.
%%%%%%%%%%%%%%%%%%%%%%%%%%%%%%%%%%%%%%%%%%%%%%%%%%%%%%%%%%%%%%%%%%%%%
\begin{acknowledgement}

This work was funded by Deutsche Forschungsgemeinschaft (DFG, German Research Foundation) under Germany's Excellence Strategy - EXC 2075 – 390740016. We acknowledge the support of the Stuttgart Center for Simulation Science (SimTech). G.M acknowledges the support of the grant RYC2022-035442-I funded by MICIU/AEI/10.13039/501100011033 and ESF+ and from  PID2024-156686NB-I00. G.M. also received support from project 20245AT016 (Proyectos Intramurales CSIC). \textbf{GM acknowledges support from ERC grant ISOCOSMOS, GA No. 101218790, funded by the European Union.} We acknowledge the computational resources provided by bwHPC and the German Research Foundation (DFG) through grant no INST \
40/575-1 FUGG (JUSTUS 2 cluster), the DRAGO computer cluster managed by SGAI-CSIC, and the Galician Supercomputing Center (CESGA). The supercomputer FinisTerrae III and its permanent data storage system have been funded by the Spanish Ministry of Science and Innovation, the Galician Government and the European Regional Development Fund (ERDF). This work is also funded by CSIC project i-LINK 23017 SENTINEL.  M. S.-N. acknowledges a Juan de la Cierva Postdoctoral Fellow proyect JDC2022-048934-I, funded by the Spanish Ministry of Science, Innovation and Universities/State Agency of Research MICIU/AEI/10.13039/501100011033 and by the European Union “NextGenerationEU”/PRTR”. V. M. R.  acknowledges support from the grant RYC2020-029387-I funded by MICIU/AEI/10.13039/501100011033 and by “ESF, Investing in your future", from the Consejo Superior de Investigaciones Cient{\'i}ficas (CSIC) and the Centro de Astrobiolog{\'i}a (CAB) through the project 20225AT015 (Proyectos intramurales especiales del CSIC), and from the grant CNS2023-144464 funded by MICIU/AEI/10.13039/501100011033 and by “European Union NextGenerationEU/PRTR”. V. M. R., and M .S.-N. acknowledge funding from grant No. PID2022-136814NB-I00 from MICIU/AEI/10.13039/501100011033 and by “ERDF, UE A way of making Europe”.

\end{acknowledgement}

%%%%%%%%%%%%%%%%%%%%%%%%%%%%%%%%%%%%%%%%%%%%%%%%%%%%%%%%%%%%%%%%%%%%%
%% The same is true for Supporting Information, which should use the
%% suppinfo environment.
%%%%%%%%%%%%%%%%%%%%%%%%%%%%%%%%%%%%%%%%%%%%%%%%%%%%%%%%%%%%%%%%%%%%%
\begin{suppinfo}

  The stationary points for the reactions on top of the ASW clusters can be retrieved at \url{10.5281/zenodo.18414194}
%This will usually read something like: ``Experimental procedures and
%characterization data for all new compounds. The class will
%automatically add a sentence pointing to the information on-line:

\end{suppinfo}

%%%%%%%%%%%%%%%%%%%%%%%%%%%%%%%%%%%%%%%%%%%%%%%%%%%%%%%%%%%%%%%%%%%%%
%% The appropriate \bibliography command should be placed here.
%% Notice that the class file automatically sets \bibliographystyle
%% and also names the section correctly.
%%%%%%%%%%%%%%%%%%%%%%%%%%%%%%%%%%%%%%%%%%%%%%%%%%%%%%%%%%%%%%%%%%%%%
\bibliography{achemso-demo}

\end{document}